\documentclass[onecolumn,useAMS,usenatbib]{mn2e}
\usepackage{graphicx}
\usepackage{natbib}
\usepackage{amssymb}

\usepackage{bm}
\usepackage{amsfonts}
\usepackage{latexsym}
\usepackage[latin1]{inputenc}
\usepackage{amsmath}
\usepackage{epsfig}
\usepackage{amsbsy}
\newcommand{\s}{\mathrm s}
\newcommand{\n}{\mathrm n}
\newcommand{\p}{\mathrm p}
\newcommand{\e}{\mathrm e}
\newcommand{\x}{\mathrm x}
\newcommand{\y}{\mathrm y}
\newcommand{\ch}{\mathrm c}

\newcommand{\veps}{\varepsilon}

\def \veps{\varepsilon}
\def\jnl@style{\it}
\def\aaref@jnl#1{{\jnl@style#1}}
\def\aaref@jnl#1{{\jnl@style#1}}

\def\aj{\aaref@jnl{AJ}}                   % Astronomical Journal
\def\apj{\aaref@jnl{ApJ}}                 % Astrophysical Journal
\def\apjl{\aaref@jnl{ApJ}}                % Astrophysical Journal, Letters
\def\apjs{\aaref@jnl{ApJS}}               % Astrophysical Journal, Supplement
\def\apss{\aaref@jnl{Ap\&SS}}             % Astrophysics and Space Science
\def\aap{\aaref@jnl{A\&A}}                % Astronomy and Astrophysics
\def\aapr{\aaref@jnl{A\&A~Rev.}}          % Astronomy and Astrophysics Reviews
\def\aaps{\aaref@jnl{A\&AS}}              % Astronomy and Astrophysics, Supplement
\def\mnras{\aaref@jnl{MNRAS}}             % Monthly Notices of the RAS
\def\prd{\aaref@jnl{Phys.~Rev.~D}}        % Physical Review D
\def\prl{\aaref@jnl{Phys.~Rev.~Lett.}}    % Physical Review Letters
\def\qjras{\aaref@jnl{QJRAS}}             % Quarterly Journal of the RAS
\def\skytel{\aaref@jnl{S\&T}}             % Sky and Telescope
\def\ssr{\aaref@jnl{Space~Sci.~Rev.}}     % Space Science Reviews
\def\zap{\aaref@jnl{ZAp}}                 % Zeitschrift fuer Astrophysik
\def\nat{\aaref@jnl{Nature}}              % Nature
\def\aplett{\aaref@jnl{Astrophys.~Lett.}} % Astrophysics Letters
\def\apspr{\aaref@jnl{Astrophys.~Space~Phys.~Res.}} % Astrophysics Space Physics Research
\def\physrep{\aaref@jnl{Phys.~Rep.}}      % Physics Reports
\def\physscr{\aaref@jnl{Phys.~Scr}}       % Physica Scripta

\title[g-modes in superfluids]{Buoyancy and g-modes in young superfluid neutron stars}

\author[Passamonti, Andersson $\&$ Ho]
{ A. Passamonti\thanks{E-mail:andrea.passamonti@oa-roma.inaf.it}$^1$,  N. Andersson$^2$ $\&$ W.C.G. Ho$^2$\\ \\
$^1$INAF-Osservatorio Astronomico di Roma, via Frascati 44, I-00040, Monteporzio Catone (Roma), Italy \\
$^2$Mathematical Sciences and STAG Research Centre, University of Southampton, Southampton SO17 1BJ, UK}

\begin{document}

%%%%%%%%%%%%%%%%%%%%%%%%%%%%%%%%%%%%  DATE  %%%%%%%%%%%%%%%%%%%%%%%%%%%%%%%%%%%%
\date{\today}

%%%%%%%%%%%%%%%%%%%%%%%%%%%%%%% PAGE RANGE  %%%%%%%%%%%%%%%%%%%%%%%%%%%%%%%%%%%%
\pagerange{\pageref{firstpage}--\pageref{lastpage}} \pubyear{}

%%%%%%%%%%%%%%%%%%%%%%%%%%%%%  MAKETITLE  %%%%%%%%%%%%%%%%%%%%%%%%%%%%%%%%%%%%%%
\maketitle

%%%%%%%%%%%%%%%%%%%%%%%%%%%%%%% FIRST PAGE  %%%%%%%%%%%%%%%%%%%%%%%%%%%%%%%%%%%%
\label{firstpage}

%%%%%%%%%%%%%%%%%%%%%%%%%%%%%  ABSTRACT  %%%%%%%%%%%%%%%%%%%%%%%%%%%%%%%%%%%%%%%

\begin{abstract}
We consider the local dynamics of a realistic neutron star core, including composition gradients, superfluidity and thermal effects. The main focus is on the gravity g-modes, which are supported by composition stratification and thermal gradients. We derive the equations that govern this problem in full detail, paying particular attention to the input that needs to be provided through the equation of state and distinguishing between normal and superfluid regions. The analysis highlights a number of key issues that should be kept in mind whenever equation of state data is compiled from nuclear physics for use in neutron star calculations. We provide explicit results for a particular stellar model and a specific nucleonic equation of state, making use of cooling simulations to show how the local wave spectrum evolves as the star ages. Our results show that the composition gradient is effectively dominated by the muons whenever they are present. When the star cools below the superfluid transition, the support for g-modes at lower densities (where there are no muons) is entirely thermal. We confirm the recent suggestion that the g-modes in this region may be unstable, but our results indicate that this instability will be weak and would only be present for a brief period of the star's life. Our analysis accounts for the presence of thermal excitations encoded in entrainment between the entropy and the superfluid component. Finally, we discuss the complete spectrum, including the normal sound waves and, in superfluid regions, the second sound. 
\end{abstract}

%%%%%%%%%%%%%%%%%%%%%%%%%%%%%  Keywords  %%%%%%%%%%%%%%%%%%%%%%%%%%%%%%%%%%%%%%%%%%%

%%%%%%%%%%%%%%%%%%%%%%%%%%%%%%% SEC. %%%%%%%%%%%%%%%%%%%%%%%%%%%%%%%%%%%%%%%%%%%
\section{Introduction} \label{sec:intro}
%%%%%%%%%%%%%%%%%%%%%%%%%%%%%%%%%%%%%%%%%%%%%%%%%%%%%%%%%%%%%%%%%%%%%%%%%%%%%%%%

It is well established from solar physics and the hugely successful Soho mission  that helioseismology \citep{2013JPhCS.440a2040G} may provide insights into the physics beneath a star's surface.  Following on from this, missions like Corot and Kepler \citep{2015arXiv150306690A} have firmly established asteroseismology as a precision science. However, in order to make use of the seismology strategy one must have a good theoretical understanding of the physics involved and the nature of the star's various oscillation modes. One also relies on nature to provide a mechanism that excites these modes to a detectable amplitude. In the case of the Sun, the interior physics is (by now) relatively well understood and we know that interior dynamics can be excited by convection. The situation is quite different when it comes to extreme objects like neutron stars. These, highly degenerate, systems also support a plethora of oscillation modes and it has been demonstrated how different aspects of supranuclear physics may influence these stars' oscillations \citep{mcd,chris}. However, we are still quite far from a complete picture (partly because the physics of the deep core of these stars remains poorly understood and partly because of difficulties in building models that account for the rich physics we know we have to include). 

The neutron-star problem is interesting (and topical) since an oscillating neutron star may radiate gravitational waves at a level that would be within reach of the advanced ground based detectors that are now coming online. In practice, this would require large-amplitude oscillations due to some kind of instability being active \citep{2011GReGr..43..409A}. Given this, it is understandable that much recent effort has been focussed on various neutron-star instabilities  (e.g. the gravitational-wave driven instability of the inertial r-modes \citep{rmode,ho2011,2014PhRvL.113y1102A}  or the f-modes \citep{2013PhRvD..87h4010P,2013PhRvD..88d4052D}). 

The matter composition in the star plays a key role in determining the fluid dynamics. Basically, the richer the ``chemistry'', the more complex the phenomenology may be. Variations in composition generally enables moving fluid elements to experience buoyancy whenever they are out of equilibrium with their surroundings. This leads to the presence of  the so-called gravity g-modes. In more familiar settings, e.g. the Sun, the main stratification is thermal. In contrast, neutron stars tend to be so cold that thermal effects are  irrelevant. Nevertheless, local fluid elements may experience buoyancy due to composition gradients \citep{rg}. This problem has been studied at some level of detail for simple neutron star models 
\citep[see, for example,][]{pass1,gk}
%(see, for example, \citet{pass1,gk}) 
and the relevance for various astrophysical scenarios, ranging from hot proto-neutron stars \citep{ferra,2011PhRvD..84d4017B} and adolescent stars \citep{chris}, to tidal interactions \citep{1995MNRAS.275..301K,2007PhRvD..75d4001F}, various instabilities \citep{1999MNRAS.307.1001L,2013ApJ...769..121W}, and the dynamics associated with core collapse and the generation of gravitational waves \citep{ott}, has been considered. More realistic neutron-star models have not yet been considered in particular detail, although it is known that interfaces associated with phase transitions will lead to the presence of a family of modes closely related to the g-modes \citep{mini}. It has also been demonstrated that the onset of superfluidity has a key influence on the buoyancy that supports the g-modes. In simple models, the support for the g-modes may, in fact, disappear altogether \citep{1995A&A...303..515L,comer}.

The traditional view is that one would not expect the convection that drives observed oscillations in main-sequence stars like the Sun to operate in a mature neutron star. A neutron  star becomes stably stratified almost immediately (after 100s or so) after birth and,  even though the composition will affect the dynamics, the various oscillation modes are expected to be convectively stable. This conclusion appears to be challenged by recent work \citep{gusk1,gusk2} that suggests that the presence of superfluidity in the neutron star core may lead to local regions becoming convectively unstable. At first sight, this result is surprising, so it seems important to establish (first of all) whether it is correct. If the presence of superfluidity can, indeed, trigger a convection phase then we will need to understand the implications this may have for (say) the star's thermal evolution and other observable phenomena. The present investigation aims to address the first of these questions. 

We consider the local dynamics of a realistic (outer) neutron star core, focussing on the buoyancy experienced by fluid elements and the associated gravity g-modes. Our model accounts for the presence of neutrons, protons, electrons and muons, and we include finite temperature effects by introducing an entropy component. However, we  assume that the dynamics we are considering are sufficiently fast that thermal conductivity can be ignored. In essence, this means that i) we can decouple the slow evolution associated with cooling, and ii) if we allow the neutrons to becomes superfluid then the entropy is locked to any normal fluid in the mixture. Since the protons are expected to form a superconductor (we neglect all electromagnetic aspects here) in the regions we are considering, the entropy will be carried by the electrons/muons. Once we have developed the detailed framework for analysing the local (plane-wave) dynamics of this model, we provide detailed results that combine an actual cooling simulation with our derived dispersion relation. This should allow us to establish whether it is realistic to expect that convectively unstable regions may, indeed, exist at some point of a neutron star's early life.

The motivation for this work is to understand i) the effect that the interior composition has on the local dynamics of the fluid in the star, and ii) how the onset of superfluidity impacts on the results. However, there are a number of crucial  technical aspects to the analysis. Our discussion highlights the microphysics information that is required to study this problem in particular, and model detailed neutron star dynamics in general. The implementation of this information may be somewhat technical (and hence it is mainly discussed in Appendices), but it is important to understand to what extent available neutron-star equation of state models provide the required information. 

Throughout the discussion, we work in a coordinate basis, expressing vector quantities in terms of their components. We denote space-time indices by lowercase italics starting from the beginning of the alphabet, $a,b,c,...$. Spatial indices are also lowercase italics, but start from $i,j,k,...$. In both cases the Einstein summation convention is assumed. The signature of the metric $g_{ab}$ is $[-,+,+,+]$, and the covariant derivative associated with this metric is denoted by $\nabla_a$. Different fluid components are distinguished by means of a constituent index; a Roman letter $\x,\y,...$. Specifically, we use $\n$ for neutrons, $\p$ of protons, $\e$ for electrons, $\mu$ for muons and $\s$ for entropy. These matter indices are not summed over when repeated.

\section{Key ingredients of the model}
%%%%%%%%%%%
\subsection{Relativistic multifluid dynamics}

As we want to allow for the presence of superfluid components, it makes sense to build our model within the general framework for relativistic multi-fluid systems 
\citep[for a review see][]{lrr-2007}.
%(for a review see \citet{lrr-2007}). 
This means that we take as our starting point the convective variational approach to relativistic fluids in which the key variables are the various fluxes $n_\x^a$, where $\x$ is a label that identifies the fluid (in the following it will be n, p, e, $\mu$ or s). In general, these fluxes do not have to be conserved, and we would have
\begin{equation}
\nabla_a n_\x^a = \gamma_\x \ ,
\label{cons}\end{equation}
where $\gamma_\x$ follows from the relevant reaction rates. The problem simplifies in two extreme limits. In the first limit, the reactions are much faster than the dynamics. This means that a moving fluid element has time to equilibrate to its surroundings and hence its composition changes accordingly during the motion. In the opposite limit, when reactions are slow, the fluid element retains its chemical identity.  In this latter case, the mismatch between the composition of the fluid element and the neighbourhood leads to buoyancy. This can either serve to push the fluid element back in the direction it came from, which leads to a stable oscillation, or push it on in the direction of travel, which leads to an instability and large-scale convection. We will assume that we are working in the slow-reaction limit, and so ignore the $\gamma_\x$ and take the individual fluxes to be conserved. 

In the text-book approach to relativistic fluid dynamics one assumes that friction locks the different species in the fluid together, and as a result they move with a common four-velocity $u^a$. However, in a maturing neutron star the neutrons will become superfluid. This means that they will not experience friction (at least not in the usual sense) and hence they can drift relative to the other components.
 In principle, the same is true for the protons, which will become superconducting. However, in order to simplify the problem one may assume that the charged components are locked electromagnetically, forming a charge-neutral conglomerate. 
 In essence, this means that we ignore the electromagnetic field (as there will be no charge currents). This assumption may not be fully justified (as neutron stars have magnetic fields and so must support charge currents), but we will nevertheless make it as it keeps the problem tractable. 

Assuming that protons, electron, muons and entropy co-move while neutrons are superfluid, we have
\begin{equation}
n_\x^{a} = n_\x   u^{a} \, , \quad \mbox{where } \x=\p, \e, \mu, \s \ ,
\end{equation}
and
\begin{equation}
n_{\n}^{a} = n_{\n}  \left( u^{a} + v^{a} \right) \, , 
\end{equation}
where  $v^a$ is the relative velocity of the neutrons ($u^a v_a = 0$). Here and in the following we assume that this relative velocity is small enough that we can ignore red-shift factors in these expressions ($v^2\ll1$). This should be true for all situations of astrophysical interest.  The number densities $n_\x$ are then (effectively) all determined by the same observer. Above the superfluid transition, the neutrons are obviously locked to the other components. 

The  momenta that are conjugate to the particle fluxes follow from an energy $\mathcal E$ (which in turn is determined by the equation of state). This leads to 
\begin{equation}
\mu^\x_a = \left( {\partial \mathcal E \over \partial n_\x^a} \right)_{n_\y^a} \ , \quad \y\neq\x  \ .
\end{equation}
In general, this allows for the presence of entrainment; an effect that describes how one fluid is dragged along as another fluid moves. In a neutron star core, the protons and neutrons will be entrained due to the strong interaction. Later, we will demonstrate that the multi-fluid formalism accounts for this effect in a very intuitive way. However, in order to keep the initial analysis tractable, we will not account for the entrainment at this point. This means that we have the momenta
\begin{equation}
 \mu_{a}^{\x} =  \mu_{\x} u_{a } \ , \qquad \x=\p, \e, \mu,\ \s \ ,
 \end{equation}
where $\mu_\x$ are the respective chemical potentials and it is worth noting that the temperature is $T=\mu_\s$. We also have, when the neutrons are superfluid, 
\begin{equation}
\mu_{a}^{\n}   =  \mu_{\n}   \left( u_{a } +   v_{a } \right) \, .
\end{equation}

In the absence of mechanisms coupling the components, the equations of motion for this system are represented by the conservation laws \eqref{cons} (although with $\gamma_\x=0$) and the individual momentum equations;                                                                      
\begin{equation}
f_\x^a =  2 n_{\x}^{a} \nabla_{ \left[   a \right.} \mu^{\x}_{   \left. b  \right]} = 0 \ , 
\end{equation}
where the square brackets denote anti-symmetrisation.
However, when some mechanism couples various components one has to account for the relevant force ($f_\x^a\neq0$). Alternatively, one can appeal to Newton's third law and work with relevant combinations of the equations (such that the equal and opposite coupling forces cancel). In the problem we consider here, the upshot of this is that it is natural to work with the total momentum equation (obtained by adding all the component equations) and the momentum equation for the superfluid neutrons (where we ignore coupling mechanisms like the vortex mediated mutual friction \citep{mendell,sidery}). The first of these equations contains the same information as the divergence of the stress-energy tensor.

Above the superfluid transition temperature one might as well work with 
\begin{equation}
\nabla_a T^{ab} = \sum_\x  f_\x^a = 0 \ ,
\end{equation}
where 
\begin{equation}
T^{ab} = (p+\veps) u^a u^b + p g^{ab} \ ,
\end{equation}
where $p$ is the pressure and $\veps$ is the energy density. In equilibrium, the equation of state for matter is given by a barotrope $p=p(\veps)$, but as soon as we account for perturbations we need to account for the detailed composition. This is an important point, which we will return to later.

Below the superfluid transition, we can still use the sum of the momentum equations (although it will take a slightly different form as there will now be explicit terms of order $v^a$). In order to describe the new degree of freedom associated with the superfluid neutrons we complement this system by $f_\n^a = 0$. 

We also have a set of thermodynamical identities. 
As we are neglecting terms of second order in $v^{a}$, the energy density and pressure obey the usual Gibbs relation (the integrated first law of thermodynamics)
\begin{equation}
 \veps + p = \sum_\x n_\x \mu_\x \, .  
\label{firstlaw}\end{equation}
From  thermodynamics principles  (cf. the definition of the conjugate momenta and the chemical potentials) we have  
\begin{equation}
 d \veps  = \sum_\x \mu_\x dn_\x   \, ,
\end{equation}
which means that
\begin{equation}
 d p  = \sum_\x n_\x d\mu_\x \,  . \label{eq:P}
\end{equation}
These relations lay the foundation for our analysis. 

%%%%%%%%%%%%%%%
\subsection{Equilibrium configurations}

Let us now turn to the problem of building a relativistic star. 
We want to consider the perturbations of a non-rotating star in dynamical and thermodynamical equilibrium (at least on the timescale of the dynamics we are considering).
That is, our background model is a spherical star  in which all fluids co-move ($v^{a } = 0$). Considering the general metric for a spherical star (taken to be fixed later, meaning that we work in the Cowling approximation\footnote{It is worth noting that we should possibly be a little bit careful here. As shown by \citet{finn}, it may be too drastic to work with a fixed metric for horizontal, low-frequency motion like that associated with the g-modes. This is worth keeping in mind should one want to carry out a quantitatively accurate analysis of the problem.});
\begin{equation}
ds^2 = - e^{2\Phi} dt^2 + e^{2\Lambda} dr^2 + r^2 \left(d\theta^2 + \sin^2 \theta d\varphi^2 \right) \ ,
\end{equation}
it is easy to see that the  
background 4-velocity and the conjugate momenta are:
\begin{equation}
u^{a } =  \left( e^{-\Phi} , \vec{0} \right)  \,  , \qquad  \mu^{\x} _{a } =  \left( - \mu^{\x} e^{\Phi} , \vec{0} \right)  \,\ .
\end{equation}
Meanwhile, the total pressure is obtained from
\begin{equation}
 p' =  - \left( p + \veps \right) \Phi'=  - \left( p + \veps \right) g  \, .
\end{equation}
Here, and in the following, primes denotes radial derivatives. We have also
introduced the gravitational acceleration, $g$, for later convenience. 
Combined with the definition of $\Lambda$ in terms of the mass enclosed within radius $r$;
\begin{equation}
m(r)= {r\over 2} \left( 1-e^{-2\Lambda}\right) \ ,
\end{equation}
and
\begin{equation}
\Phi' = {e^{2\Lambda}\over r^2} \left( m+4\pi pr^2\right)\ ,
\end{equation}
 these are the familiar Tolman-Oppenheimer-Volkoff equations. 
For later convenience, it 
 is also worth noting that the Einstein equations lead to 
\begin{equation}
\Lambda' + \Phi ' = {r e^{2\Lambda} \over 2} \left ( p + \varepsilon\right) \ .
\end{equation}

Finally, if the neutrons are superfluid, then we also have \citep{gusa}
\begin{equation}
 \mu_\n^\prime =  - \mu_{\n}    \Phi' \quad  \longrightarrow \quad \tilde\mu_\n \equiv \mu_\n e^{\Phi} = \mathrm{constant}  \, .  \label{eq:back}
\end{equation}

%\comment{FIGURE: Mass vs radius for BSk20, showing maximum mass etc.}

%%%%%%%%%%%%%% FIG %%%%%%%%%%%%%%%%%%
%------------------------------FIG. 1------------------------------------------%
\begin{figure*}
\begin{center}
\includegraphics[height=75mm]{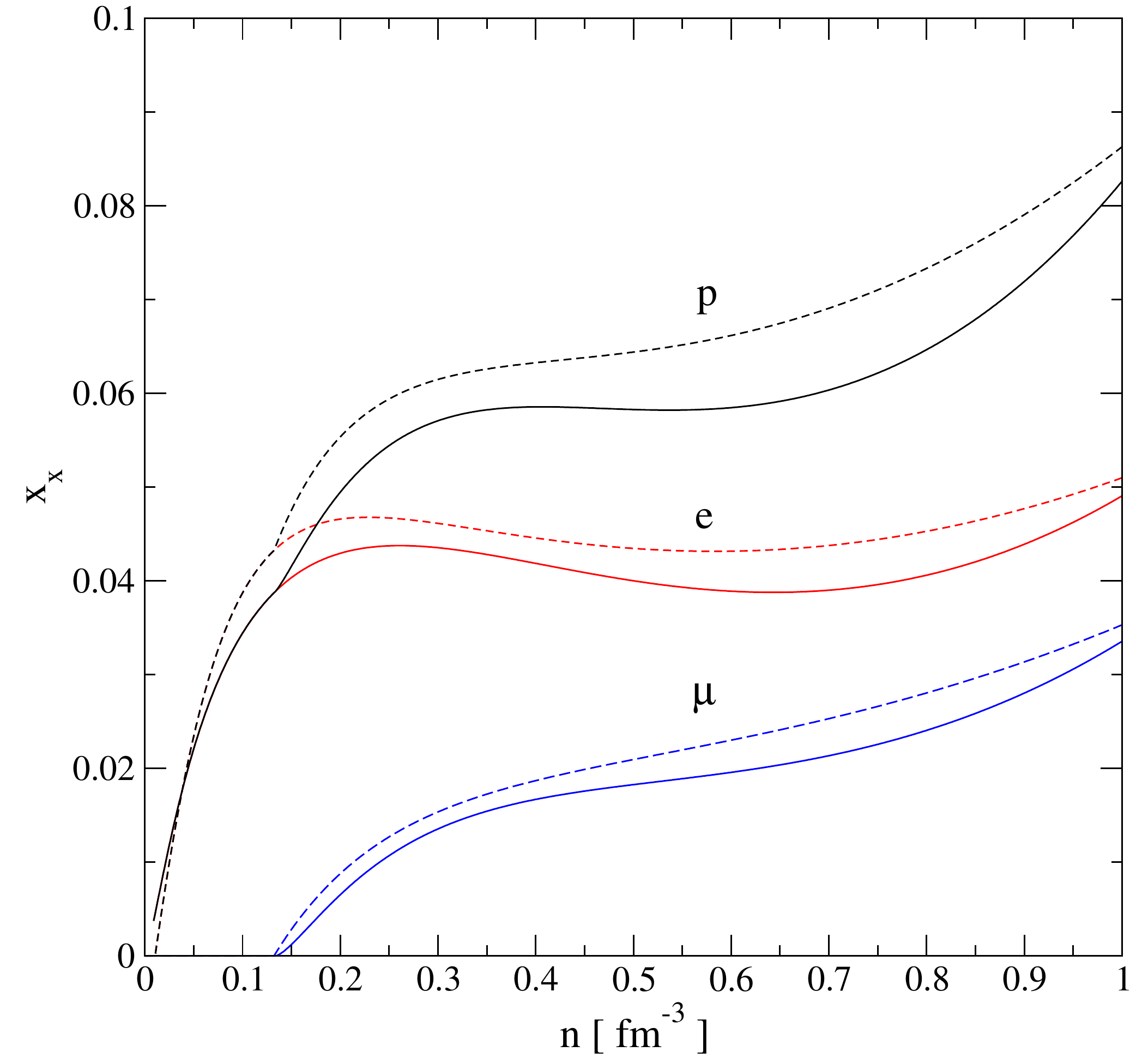} 
\includegraphics[height=75mm]{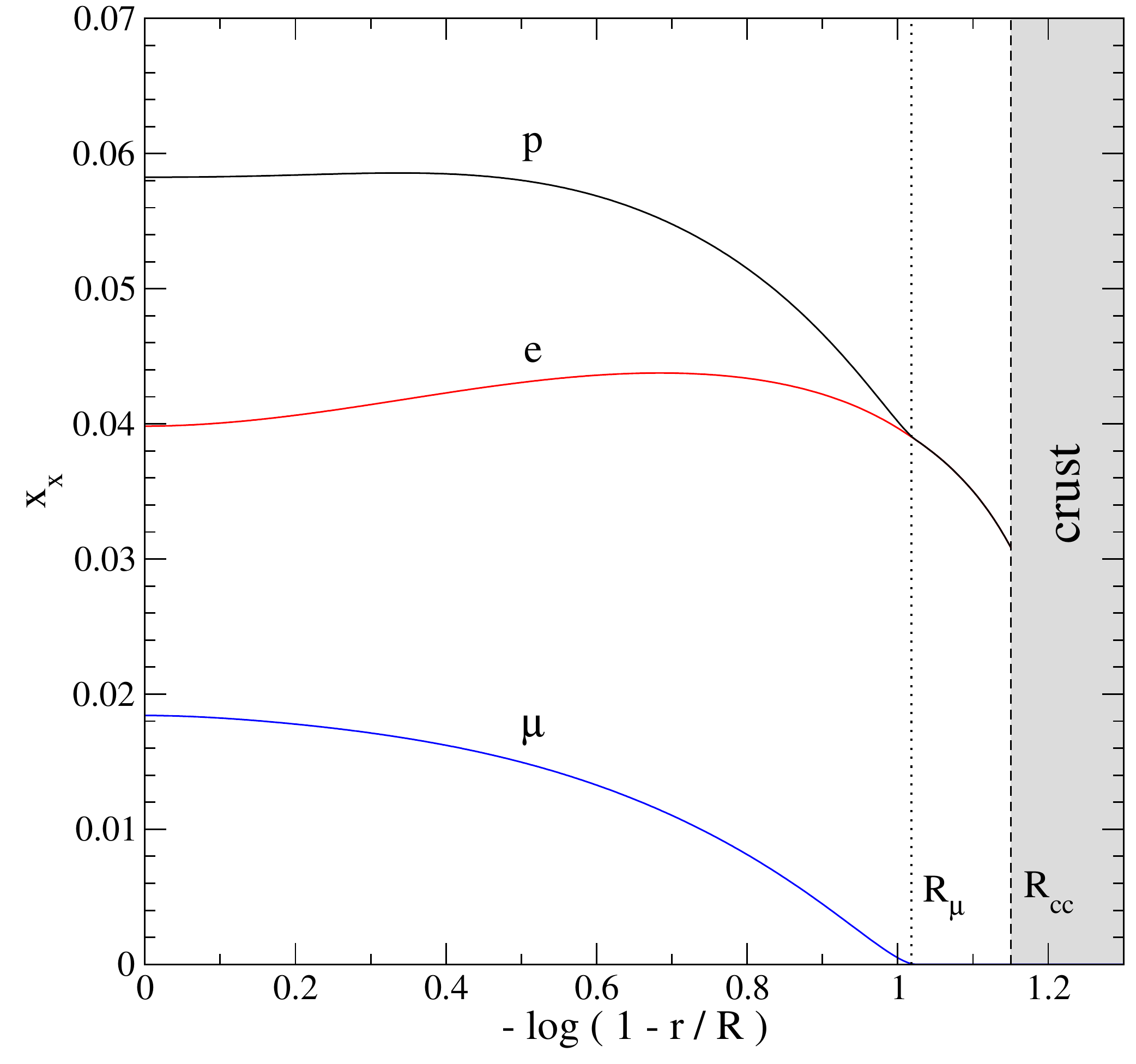} 
\caption{  Composition gradients for our chosen equations of state  \citep[BSk20 from][]{bsk2} 
%(BSk20 from \citet{bsk2}) 
and the particular stellar model we focus on throughout the paper. The left-hand panel shows the  fractions of protons, electrons and muons 
versus the total baryon density, $n$, and compares the approximate description used in this work (solid lines) with the analytical fits to the full numerical solution obtained by \citet{pote} (dashed lines). 
The right-hand panel shows the core composition for our stellar model, which has mass $M=1.4 M_{\odot}$ and radius $R=11.66$~km. The  vertical lines in the right-hand panel represent, respectively,  
the position where $x_{\mu} = 0$ ($R_\mu=10.53$~km) and the crust/core interface ($R_{cc}=10.82$~km). Note that the right-hand panel is using a stretched radial coordinate for which the star's surface is pushed to infinity. Hence, the crust region (indicated as grey in the figure) has been truncated, but this is unimportant as we do not consider the dynamics of that region. 
\label{fractions}}
\end{center}
\end{figure*}
%------------------------------------------------------------------------------%

%\begin{figure}
%\begin{center}
%\includegraphics[height=7cm]{fractions.eps}
%\end{center}
%\caption{Particle fractions for the BSk20 model. }
%\label{fractions}
%\end{figure}
%%%%%%%%%%%%%%%%%%%%%%%%%%%%%%%%%%%

%%%%%%%%%%
\subsection{Composition gradients}

In order to complete the background model and discuss the internal composition of a given star, we obviously need to provide an equation of state. In principle, we can assume that this is given in terms of an energy (density) $\veps = \veps(n,x_\x)$, where $n=n_\n+n_\p$ is the total baryon number density and $x_\x= n_\x/n$ are the various species fractions. Once we have this relation, we can readily work out the pressure from \eqref{firstlaw}. There are numerous proposed ``realistic'' equations of state available in the literature, so one might think we would be spoilt for choice. Unfortunately (or perhaps fortunately?) this is not the case. The problem is that most tabulated equations of state do not provide all the information we need in order to study neutron star oscillations. Basically, the tabulated data often assumes that the matter is in equilibrium, whereas we need to be able to track how the matter responds when driven out of equilibrium by the fluid motion. The equilibrium configuration must satisfy three conditions;
\begin{equation}
 \mbox {beta equilibrium:} \qquad \mu_\n = \mu_\p+\mu_\e \ , 
\label{one}\end{equation}
\begin{equation}
\mbox{lepton balance:} \qquad \mu_\mu=\mu_\e \ , 
\label{two}\end{equation}
and
\begin{equation}
\mbox{charge neutrality:} \qquad n_\p = n_\e + n_\mu \ , 
\label{three}\end{equation}
but the perturbations may be such that a moving fluid element violates one or more of these conditions. In fact, it is precisely this mismatch that leads to the buoyancy that gives rise to the presence of g-modes. In other words, we need to get the physics input from a general energy functional that allows us to model matter out of equilibrium. For practical reasons, it would be advantageous if the equation of state was given in a parameterised analytic form rather than as a numerical table. We will need to work out various partial derivatives and it would be an extremely tricky exercise to do this in a thermodynamically consistent way starting from tabulated data (this should be clear from the discussion in Appendix A). For these reasons, we base our analysis on the family of equations of state provided by the Brussels-Montreal collaboration \citep{bsk1,bsk2,bsk3}. Even though we focus on one particular family of models, the parameter space is still overwhelming. Hence, we will provide actual results only for a single stellar model. The star we consider is constructed from the BSk20 equation of state \citep{bsk2}.  A discussion of the general features of this equation of state model and stars built from it can be found in \citet{bsk3} and \citet{pote}. Our particular stellar model has mass $M=1.4 M_{\odot}$ and radius $R=11.66$~km. This should be a fairly ``average'' neutron star.

Let us now consider the interior composition of this model. 
In principle, this problem must be solved numerically. However, in the spirit of the present study it would be preferable to have an analytic approximation\footnote{Note that the analytic fits for the particle fractions from \citet{pote} are not directly useful for the present problem as they assume matter at equilibrium and we need to quantify the response to deviations from this equilibrium state.}. Fortunately, such an approximation is readily obtained if we note that the symmetry energy  $S(n)$ (which represents the energy cost in replacing protons with neutrons in symmetric nuclear matter) satisfies 
\citep[see][]{chamel08}
%(see \citet{chamel08})
\begin{equation}
\mu_\e \approx \hbar c \left( 3\pi^2 n x_\e\right)^{1/3} \approx 4 S(n) (1 - 2x_\p) \ ,
\end{equation}
where $\hbar$ is the reduced Planck's constant and $c$ is the speed of light (which is taken to be unity throughout most of the discussion of dynamics, but reinstated in parts of the discussion of the equation of state input), and we assume that the electrons are sufficiently relativistic that we can neglect their rest mass. This is a good approximation for a neutron star core (and most of the crust). For a typical neutron star, the outer layers of the core only contains neutrons, protons and electrons. Then we can use the condition for charge neutrality to replace $x_\p$ with $x_\e$ on the right-hand side of the equation. Thus we obtain
\begin{equation}
x_\e \approx {1 \over 3\pi^2 n} \left( {4S \over \hbar c} \right)^3 \left[ 1 + {2 \over \pi^2 n} \left( {4S \over \hbar c} \right)^3\right]^{-1}\ ,
\label{xeapp}\end{equation}
Once the muons appear, the problem becomes a little bit more involved, but in reality \eqref{xeapp} remains quite accurate (basically since the fraction of charged particles is small). This means that we can solve \eqref{two} for the muon fraction. To do this, we need 
\begin{equation}
\mu_\mu = m_\mu c^2 \left( 1 + {\hbar^2 k_{F\mu} \over m_\mu^2 c^2} \right)^{1/2}\ ,
\end{equation}
where $k_{F\mu}=\left( 3 \pi^2 n x_\mu\right)^{2/3}$ is the Fermi wavenumber for the muons. Thus we arrive at 
\begin{equation}
x_\mu \approx {1 \over 3 \pi^2 n} \left( {m_\mu c \over \hbar} \right)^3 \left[ \left( {\hbar  \over m_\mu c} \right)^2 \left( 3 \pi^2 n\right)^{2/3} - 1\right]^{3/2}\ .
\end{equation}
Finally, the proton fraction follows from $x_\p = x_\e+x_\mu$. As an illustration, the resulting fractions for the BSk20 model are shown in the left panel of Figure~\ref{fractions}. These results agree well with the parameterised expressions given by \citet{pote}, which are based on the numerical solution to the problem. Our approximate fractions are accurate enough for our purposes. The right-hand panel of Figure~\ref{fractions} shows the composition of our model star, indicating the crust-core transition ($R_{cc}=10.82$~km here) as well as the radius at which the muons first appear ($R_\mu=10.53$~km in our case). 

%%%%%%%%%%
\subsection{Thermal gradients}

The buoyancy of a local fluid element depends not only on the matter composition, it can also be affected by thermal gradients. In order to account for this aspect, which should be important during the early stages of a neutron star's life, we carry out a detailed cooling simulation for our model star.

\begin{figure*}
\begin{center}
\includegraphics[height=75mm]{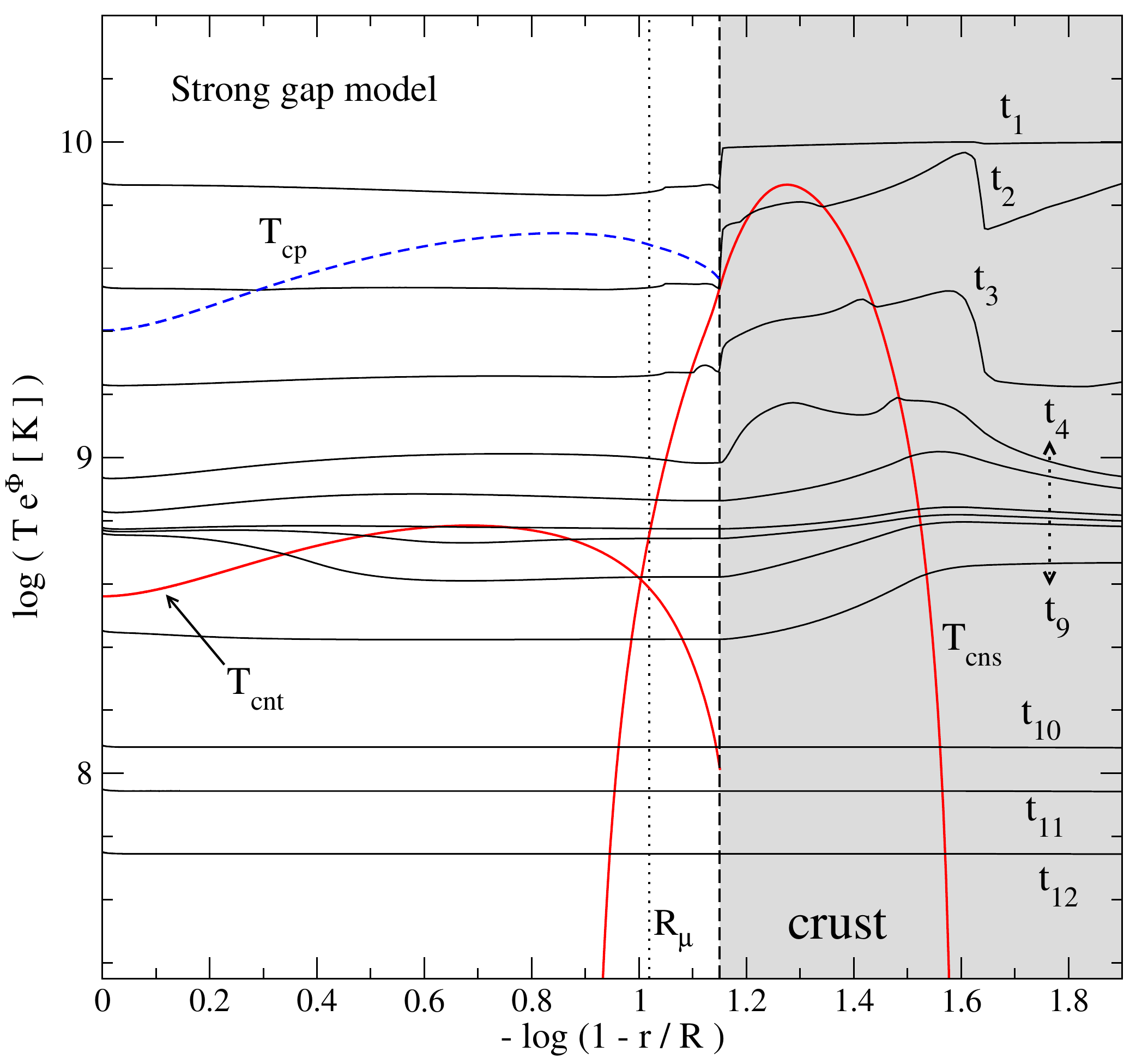} 
\caption{A sequence of cooling results for particular choices of the superfluid pairing gaps. The initial temperature for our evolution is taken to be a constant $Te^\Phi=10^{10}\mbox{ K}$. We show the critical temperatures for the onset of neutron
superfluidity in the singlet state in the crust (ns; red solid line) and triplet
state in the core (nt; red solid line) and proton superconductivity in the core
(p; blue dashed line). Specifically, the neutron singlet model is that from \citet{1989PhLB..222..173A},
the neutron triplet model is taken from \citet{1985NuPhA.442..163A},
and we use the proton superconductivity model from \citet{1993NuPhA.555...59C}. The parametrisation of the gaps is the same as in \citet{hoetal15}. Note that the critical temperature for neutron superfluidity in the core is relatively high compared to many other models in the literature. We have considered more moderate cases, but focus on this case because it was the only model that led to the presence of a brief era of unstable g-modes (see Figure~\ref{instab} later). Note also that thermal gradients are more pronounced  in the crust. The results in this figure provide input for our discussion of the thermal effects  for particular ages of the star ($t_i$ in the figure).
\label{fig:cooling}}
\end{center}
\end{figure*}

The temperature evolution of an isolated neutron star is determined by
the relativistic equations of energy balance and heat flux 
\citetext{see \citealp{2004ARA&A..42..169Y} and \citealp{2006NuPhA.777..497P}, for reviews}.
%(see \citet{2004ARA&A..42..169Y} and \citet{2006NuPhA.777..497P}, for reviews).
 We solve this system using the
method described in \cite{hoetal12}. The physics inputs for these equations are the heat capacity, emissivity of
neutrinos (which is the primary cooling process at ages
$\lesssim 10^6\mbox{ yr}$), and thermal conductivity.
The heat capacity is the sum of contributions from the constituent particles, also given in \cite{hoetal12}.
Various neutrino emission processes contribute to the total 
emissivity, with the primary processes being the modified Urca mechanism and
Cooper pair formation and breaking 
\citetext{see \citealp{hoetal12}, and references therein}; 
%(see \cite{hoetal12}, and references therein); 
the BSk20 EoS does not produce neutron stars that undergo neutrino
emission via the direct Urca mechanism.
Thermal conductivities in the core are calculated using results from
references given in \cite{hoetal12}, while those in the crust are calculated
using {\small CONDUCT13}\footnote{http://www.ioffe.ru/astro/conduct/}.
Note that, due to high thermal conductivity the crust and the core are roughly isothermal after $\sim 10-100\mbox{ yr}$ (see Fig.~\ref{fig:cooling}), although there are still slight variations at later times due to the superfluidity.
Two other important factors affecting neutron star cooling behaviour are the
composition of the outer layers (envelope) of the neutron star crust and
whether the stellar interior is superfluid and/or superconducting.  For the
former, the envelope serves as a heat blanket, with envelopes composed of
light elements conducting heat more efficiently than heavy element envelopes.
For simplicity,  we only consider neutron stars with an iron envelope.
For the second factor, superfluidity and superconductivity suppress heat
capacities and some neutrino emission mechanisms, as well as enhancing
neutrino emission through the Cooper pairing process.

The initial temperature for our evolutions is taken to be a constant $Te^\Phi=10^{10}\mbox{ K}$.
Figure~\ref{fig:cooling} shows the time evolution of temperature at different ages of our model star.
Also shown are the critical temperatures for the onset of neutron
superfluidity in the singlet state in the crust and triplet
state in the core  and proton superconductivity in the core. 
Specifically, we use the neutron singlet model from \citet{1989PhLB..222..173A},
the neutron triplet model is taken from \citet{1985NuPhA.442..163A},
and we use the proton superconductivity model from \citet{1993NuPhA.555...59C}. Critical temperatures are calculated using parametrised gap models
following the prescription from \citet{hoetal15}.
Note that, as we do not yet have self-consistent calculations of the relevant pairing gaps for neutrons and protons for the BSk equations of state (or, indeed, any other proposed equation of state), we are ``forced'' to use this phenomenological prescription. Note that, in the following we do not distinguish between singlet and triplet pairing for the neutron superfluid. In essence, we take the critical temperature to be $T_{c\n}=\mathrm{max}(T_{c\n s},T_{c\n t})$. We also assume that $T_{c\p}> T_{c\n}$ throughout the core (as in Figure~\ref{fig:cooling}).

Our results show that,  at very early times, the core cools more rapidly than the crust
via stronger neutrino emission, so that the crust is generally at higher
temperatures.  A cooling wave travels from the core to the surface, eventually bringing
the neutron star to a relaxed, isothermal state.
We also see the effect of superfluidity, i.e., faster cooling after neutrons
become superfluid in the crust at early times and in the core at later times. These effects are strongest in regions near the critical temperatures, in accordance with the expectations from previous work \citep{2004A&A...423.1063G}. 

%------------------------------FIG. 2------------------------------------------%
%------------------------------------------------------------------------------%

\subsection{Accounting for thermal effects}

Given the results of the recent analysis by \citet{gusk1} and the suggestion that thermal effects may lead to convective regions in a cooling neutron star, we need to be able to consider both superfluidity and thermal effects. A key feature of this problem is that the leptons (electrons and muons) are the main carriers of entropy in a superfluid neutron star core. 
As long as the local temperature is far below the critical temperature for the onset of superfluidity, it is safe to ignore the presence of thermal excitations in the superfluid. However, these excitations play an important role near the transition. In essence, there may always be a local region (near the critical density for the given temperature) where excitations must be accounted for. As discussed by \citet{and13}, this effect can be quantified in terms of entrainment between the superfluid and the normal component (entropy). In practice, this leads to the superfluid component locking to the normal component at the transition. A key question concerns to what extent the superfluid transition is local. If it is, then one may be able to avoid a detailed analysis and simply impose suitable junction conditions. If, on the other hand, the interface is extended over a macroscopic region, then any study of neutron star dynamics will have to account for the detailed transition. As we will account for the presence of thermal excitations, our analysis should shed some light on this issue.

Our aim is to combine the results of the cooling simulations with a local analysis of the fluid dynamics. When it comes to the inclusion of thermal effects, the discussion involves approximations and it is useful to explain what they are. First of all, the cooling simulation is carried out for a stellar model with fixed size, internal composition and so on, by solving the heat equation on top of a passive stellar configuration. This approach is standard \citep{2004ARA&A..42..169Y,2006NuPhA.777..497P} and should be adequate as long as one can neglect the effects of thermal pressure compared to the cold degeneracy pressure. As our cooling simulation starts at a temperature of $10^{10}$~K~$\sim$1~MeV and the typical chemical potentials are of order 100~MeV, this approximation should be safe. The errors involved are certainly smaller than our level of ignorance regarding the supranuclear physics.

When it comes to quantifying the thermal effects, it is useful to consider the following argument: The total pressure has to satisfy
\begin{equation}
p' =  \sum_{\x=\n,\p,\e,\mu} n_\x \mu_\x^\prime  + s T' = - \left( \sum_{\x=\n,\p,\e,\mu} n_\x \mu_\x + sT \right)  \Phi'  =  - \left(  p +\veps \right) \Phi'\ ,
\end{equation}
where we assume that the star is composed of neutrons, protons, electrons and muons (as before) and the entropy density is given by $s$. In a region where the neutrons are superfluid, they must satisfy \eqref{eq:back}. If we also impose the equilibrium conditions (balance between electrons and muons and charge neutrality)  we are left with a relation that can be written;
\begin{equation}
n_\p {d\over dr} \left[ (\mu_\p+\mu_\e) e^\Phi\right] = - s {d\over dr} \left(T e^\Phi \right) = - s {dT^\infty \over dr} \ .
\end{equation}
Combining this with  \eqref{eq:back} we arrive at
\begin{equation}
 {d\over dr} \left[ \left(\mu_\p+\mu_\e-\mu_\n\right) e^\Phi \right] =  - {s\over n_\p} {dT^\infty \over dr} \ .
\end{equation}
This relation 
\citetext{also given by \citealp{gusa}} 
%(also given by \citet{gusa}) 
tells us that a model that satisfies the various chemical equilibrium conditions must also be in thermal equilibrium. That is, we must have 
$T^\infty=$~constant.
This is, of course, not going to be true during the early cooling phase, see Figure~\ref{fig:cooling}. In order to make the model consistent we would need to add the thermal contributions to the relevant chemical potentials. This would then lead to a gradual evolution of the star's composition during the cooling; essentially the tail-end of the de-leptonisation that dominates the first 100~s or so of the star's life \citep{burrows}. However, as this effect is very small during the phase that we are considering it should be safe to neglect it. 

In quantifying the fluid dynamics we do, however, want to account for thermal effects. We do this by considering the equation of state as a zero-temperature model with thermal contributions added perturbatively. Assuming that only the electrons and muons contribute and treating each as a relativistic Fermi gas \citep{prakash}, we have $s=s_\e+s_\mu$ where
\begin{equation}
s_\x =  {\pi^2 \over 2} {k_B T n_\x\over T_{F\x}} \ , \quad \x=\e,\mu\ ,
\label{xs1}\end{equation}     
and the relevant Fermi temperature is given by
\begin{equation}
k_B T_{F\x} = \hbar c \left( 3\pi^2 n_\x\right)^{1/3} \ ,
\end{equation}
(note that, for simplicity, we are ignoring the muon mass here). 
``Integrating'' the definition of the temperature (as the entropy chemical potential);
\begin{equation}
T = \left( {\partial \veps \over \partial s}\right)_{n_\x}\ ,
\end{equation}
we see that the thermal energy is
\begin{equation}
\veps_{th} = {\pi^2 \over 4}  k_BT \left({n_\e T\over T_{F\e}}+{n_\mu T\over T_{F\mu}}\right) \ ,
\end{equation}
and the thermal pressure, which follows from the integrated first law \eqref{firstlaw} , is
\begin{equation}
p_{th} =  {\pi^2 \over 12}  k_B T \left({n_\e T\over T_{F\e}}+{n_\mu T\over T_{F\mu}}\right) \ .
\end{equation}
This contribution tends to be (very) small compared to the cold degeneracy pressure, so we only include it  whenever it plays a leading role in the analysis.

In the case of a superfluid neutron star core, we will work with $S=s/n_\e$ (cf. Equation~\eqref{entper}). For this variable, the above relations imply that we have
\begin{equation}
S' = S\left[ {T'\over T} - {n_\e'\over n_\e} + {2\over 3}\left( {n_\e'+(n_\e/n_\mu)^{1/3}n_\mu' \over n_\e + (n_\e n_\mu^2)^{1/3} }\right)  \right] \ . \label{eq:dSdr}
\end{equation}
In the low-density region of the core where the muons are yet to appear, this reduces to
\begin{equation}
S' = S\left[ {T'\over T} - {n_\e'\over 3n_\e}  \right] \ .
\end{equation}

%%%%%%%%%%%%%%%%%%%%%%%%%%%%            
\section{Composition g-modes of a cold normal-fluid star}

In order to establish our approach to the problem, and provide  useful results for comparison, let us begin our  analysis 
by revisiting the standard g-mode problem \citep{rg}, ignoring superfluidity and thermal effects. That is, we consider a stratified star where all fluid constituents move together, in such a way that there is only one fluid four velocity to worry about. In this case the equations that govern deviation from a static equilibrium are (i) the perturbed Euler equation
\begin{equation}
\left( \veps + p \right) \partial_{t}   \delta u_{k}  = - e^{\Phi}  \partial_{k} \delta p   -  e^{\Phi} \left(  \delta \veps + \delta p \right) \partial_{k} \Phi  \ , \label{momtot}
\end{equation}
and (ii) the baryon number conservation law
\begin{equation}
e^{-\Phi}  \partial_{t}   \delta n  + \partial_{k} \left(  n  \delta u^{k} \right)
+ n\left[ \left(\Lambda' + \Phi '   + \frac{2}{r} \right)    \delta u^{r} 
+ \cot \theta  \,   \delta u^{\theta}  \right]  = 0 \, . 
\label{constot}
\end{equation} 
The system of equations is closed once we specify an equation of state. Looking ahead to the problem of main interest, that of a superfluid neutron star core, we assume that the star is composed of a mixture of neutrons, protons, electrons and muons.  Our aim is to establish how variations in composition leads to buoyancy and the emergence of a set of g-modes. 

%%%%%%%%%%%%%%%%%%%%%%%%
\subsection{The dispersion relation}

We assume that the dynamics is fast enough that the composition of each fluid element can be considered frozen\footnote{The veracity of this assumption can always be checked a posteriori by comparing the inferred dynamical timescales to the relevant dissipative timescales, etcetera.}. This is equivalent to stating that the Lagrangian variation ($\Delta$) of the particle fractions $x_\x$ vanish\footnote{For later reference it is worth noting how this condition is derived. The conclusion follows from the general result for the Lagrangian perturbation of each number density 
$$
\Delta n_\x = - {n_\x \over 2} \perp^{ab}\left[ \delta g_{ab} + 2\nabla_{(a}\xi_{b)}\right] \ .
$$
Whatever the fluid motion is, the metric variation $\delta g_{ab}$ is the same for all components. If two components are locked, in the sense that their displacements are the same, it is easy to show that the Lagrangian perturbation of the ratio of the two number densities must vanish.}. This means that we have
\begin{equation}
\Delta x_\x = 0 \ , \qquad \longrightarrow \qquad \delta x_\x = - \xi^r x_\x^\prime \ ,
\label{Lagdef}\end{equation}
where $\delta$ denotes Eulerian perturbations and $\xi^a$ is the Lagrangian displacement vector. In the case we are considering (a static background star) we have
\begin{equation}
\delta u^{a } =  \mathcal{L}_{u} \xi^{a} =  e^{-\Phi} \partial_{t} \xi^{a} \ .
\end{equation}
where $\mathcal L_u$ represents the Lie derivative along the background four velocity.

In order to study the dynamics of the star, we  assume a harmonic  time-dependence $e^{i\sigma t}$ and use the standard (polar) decomposition of the displacement vector in terms of spherical harmonics $Y_l^m(\theta,\phi)$. We then have (suppressing sums over $l$ and $m$, since the different multipoles decouple for a spherical background star) 
\begin{align}
&\xi^{r} =   W  \, Y_l^m  e^{i\sigma t} \, , \label{Wdef} \\
&\xi^{\theta} = \frac{V}{ r^2}  \,  \partial_\theta Y_l^m  e^{i\sigma t}  \label{Vdef1}\, , \\
&\xi^{\phi} = \frac{V}{ r^2 \sin \theta ^2}  \,  \partial_\phi Y_l^m  e^{i\sigma t} \label{Vdef2} \, ,
\end{align}
where $W$ and $V$ are functions of $r$.
Scalar quantities, like the perturbed pressure, are also expanded in spherical harmonics, which means that we have $\delta p = \hat p Y_l^m e^{i\sigma t} $ etcetera. In the following, hats are frequently used to denote amplitudes of perturbed quantities. These variables are functions of $r$ only.

In order to make progress on the problem, we need to decide which variables to work with. In this first example, we will choose to work with the perturbed pressure, $\delta p$, and the radial component of the displacement vector, $W$. This means that we need to obtain relations for the perturbed number density, $\delta n$, and the perturbed energy density, $\delta \veps$, from the equation of state. As we are assuming that the composition of each fluid element is frozen on the timescale of the motion, it is natural to consider the equation of state as a function\footnote{Note that this is \underline{not} the form we assumed when we discussed the equilibrium configuration. We are using the pressure, $p$, rather than baryon number density, $n$, as one of the primary variables. This is not necessarily the most practical choice, but it facilitates a direct comparison with the work of \citet{gusk1} and \citet{gusk2}} $\veps = \veps(p, x_\x)$. Moreover, it is convenient to work with the enthalpy, $ w = p + \veps$. This means that we have
\begin{equation}
\delta w = \left[ 1 + \left( {\partial \veps \over \partial p}\right)_{x_\x} \right] \delta p + \sum_{\x} \left( {\partial \veps \over \partial x_\x}\right)_p \delta x_\x \ .
\label{delw}\end{equation}
Here, and throughout the rest of this Section, the sum over $\x$ involves $\p, \e$ and $\mu$.
If we define the speed of sound as
\begin{equation}
c_s^2 = \left( {\partial p \over \partial \veps}\right)_{x_\x} \ ,
\end{equation}
and use \eqref{Lagdef} we see that \eqref{delw} leads to
\begin{equation}
\hat w = \left( 1 + {1\over c_s^2} \right) \hat p - \left(  \sum_{\x}  {\veps \over \Gamma_\x} x_\x^\prime\right) W \ ,
\end{equation}
where we have defined (for later convenience)
\begin{equation}
{1\over \Gamma_\x}  = {1\over \veps}  \left( {\partial \veps \over \partial x_\x}\right)_{p,x_\y}\ , \qquad \y \neq \x \ .
\label{Gammax}\end{equation}
Finally introducing
\begin{equation}
N_c^2 = - g \sum_{\x} {x_\x^\prime \over \Gamma_\x}  \ ,
\end{equation}
we arrive at
\begin{equation}
\hat w = \left( 1 + {1\over c_s^2} \right) \hat p +  {\veps \over g} N_c^2 W \ .
\end{equation}

With these various definitions, we also have the radial component of the Euler equation;
\begin{equation}
w \sigma^2 W e^{-2(\Phi-\Lambda)} = \hat p^\prime + g \hat w \ , 
\end{equation}
and the angular component;
\begin{equation}
w \sigma^2 V e^{-2\Phi} = \hat p \ . 
\end{equation}
The latter immediately allows us to eliminate $V$ from the analysis. Finally, the baryon number conservation law leads to 
\begin{equation}
W^\prime + \left( {n^\prime \over n} + {2\over r} + \Lambda^\prime\right) W = {l(l+1) \over r^2}V - {\hat n \over n} \ .
\end{equation}
Here we see that we  need
\begin{equation}
{\hat n \over n} = {1\over \Gamma_1^b} {\hat p \over p} + \sum_{x} {1\over \Gamma_\x^b} \delta x_\x \ ,
\end{equation}
where 
\begin{equation}
{1\over \Gamma_1^b} = {p\over n} \left( {\partial n \over \partial p}\right)_{x_\x} \ , 
\end{equation}
(the index $b$ indicates that the adiabatic index is associated with the baryon number $n$) and
\begin{equation}
 {1\over \Gamma_\x^b} = {1\over n} \left({ \partial n \over \partial x_\x}\right)_{p,x_\y}\ , \qquad \y \neq \x \ .
\end{equation}
Defining
\begin{equation}
A_b^\x = {x_\x^\prime \over \Gamma_\x^b} \ ,
\end{equation}
we have
\begin{equation}
{\hat n \over n} = {\hat p \over p \Gamma_1^b}  - W \mathcal A_b \ ,
\end{equation}
where $\mathcal A_b = \sum_\x A_b^\x $.
Introducing also
\begin{equation}
\mathcal L_b^2 = \frac{p}{w} {l(l+1) \over r^2} e^{2\Phi} \Gamma_1^b \ ,
\end{equation}
we have the two equations
\begin{equation}
W^\prime + \left( {n^\prime \over n} + {2\over r} + \Lambda^\prime - \mathcal A_b \right) W = \left( {\mathcal L_b^2 \over \sigma^2} - 1\right) {\hat p \over p\Gamma_1^b}  \ ,
\label{Weqn}\end{equation}
and
\begin{equation}
\hat p^\prime + g \left( 1 + {1\over c_s^2} \right) \hat p  = e^{-2(\Phi-\Lambda)} w \left[ \sigma^2 - \mathcal N_c^2 \right]  W \ ,
\label{peqn}\end{equation}
where
\begin{equation}
\mathcal N_c^2 = \frac{\veps}{w} e^{2(\Phi-\Lambda)} N_c^2 \ .
\label{normbrunt}\end{equation}

In order to analyse the local dynamics of the problem we now make the plane-wave assumption, i.e.  assume that all perturbations depend on position  as $e^{ikr}$ where $kr \gg 1$, as the focus is on the local dynamics. We also note that we can introduce integrating factors on the left-hand side of each equation\footnote{For completeness; the integrating factors we need are, for Equation \eqref{Weqn};
\begin{equation}
 nr^2 \exp \left[ \Lambda - \int \mathcal A_b dr \right] \ ,
\end{equation}
and for Equation \eqref{peqn};
\begin{equation}
 \exp\left[ \Phi+ \int {g/c_s^2} dr \right] \ .
\end{equation}}. However, it is straightforward to demonstrate 
\citetext{see \citealp{comer}} 
%(see \citet{comer}) 
that these integrating factors do not affect the final dispersion relation. The upshot of this is that we can effectively ``ignore'' the second term of each equation. 
Thus, we arrive at the two relations (ignoring the overall $e^{ikr}$ factor, as this should not cause any confusion)
\begin{equation}
ik W = { \mathcal L_b^2 - \sigma^2 \over \Gamma_1^b \sigma^2 } {\hat p \over p} r^2 e^{\Lambda - \Phi} \ ,
\end{equation}
\begin{equation}
ik \hat p = e^{-(\Phi-\Lambda)} w \left( \sigma^2 - \mathcal N_c^2\right) \frac{W}{r^2}  \ .
\end{equation}

Combining these  we have 
\begin{equation}
{w \over p} \left( {\mathcal L_b^2-\sigma^2 \over \Gamma_1^b \sigma^2}\right) \left( \sigma^2 -\mathcal N_c^2\right) + \tilde k ^2 = 0  \ ,
\label{littledisp}\end{equation}
where
\begin{equation}
\tilde k= e^{\Phi-\Lambda} k \ .
\end{equation}

This should be the complete dispersion relation, without approximations other than the plane-wave assumption. We see that there are two sets of solutions; representing travelling waves in two limits. As we need $k^2>0$, we must have either $\sigma^2>\mathcal L_b^2 > \mathcal N_c^2$ (generally expecting $\mathcal N_c^2 \ll \mathcal L_b^2$) or  $\sigma^2<  \mathcal N_c^2< \mathcal L_b^2 $.  The first set of solutions correspond to high-frequency sound waves given by;  
\begin{equation}
\sigma^2 \approx \left[ {l(l+1) \over r^2} + k^2 e^{-2\Lambda} \right] {p \Gamma_1^b e^{2\Phi}\over w}  \approx \frac{p}{w}  {l(l+1) \over r^2} e^{2\Phi} \Gamma_1^b =\mathcal L_b^2 =  {l(l+1) \over r^2} e^{2\Phi} {n\over w} \left( {\partial p \over \partial n}\right)_{x_\x}  \ .
\label{sfreq}
\end{equation}
The second approximation is valid whenever $kr\ll l$, an assumption we will use to simplify the multi-fluid problem later. 
Meanwhile, for low frequencies we have  (composition) g-modes with  frequency;
\begin{equation}
\sigma^2 \approx \left[ 1 - { k^2 r^2 e^{-2\Lambda} \over l(l+1)} \right] \mathcal N_c^2 \approx \mathcal N_c^2 =- g \frac{\veps }{w} e^{2(\Phi-\Lambda)}  \sum_{\x=\p,\e,\mu}  \frac{ x_{\x}^{\prime} }{\Gamma_{\x}}  \ .
\label{gfreq}\end{equation}
This result shows how the gradients of the different particle fractions provide the buoyancy that leads to the g-modes. We also see that the modes will be unstable if the leading gradient increases outwards in the star (provided $\Gamma_\x>0$). This is the familiar criterion for the onset of convection. Of course, considering the actual composition gradients from Figure~\ref{fractions} we see why convection is not expected to take place in cold neutron stars. The composition stratification is stable (although it is worth noting that the there is a region where the electron contribution has a destabilising effect in this particular example).

%------------------------------FIG. 3------------------------------------------%
\begin{figure*}
\begin{center}
\includegraphics[height=75mm]{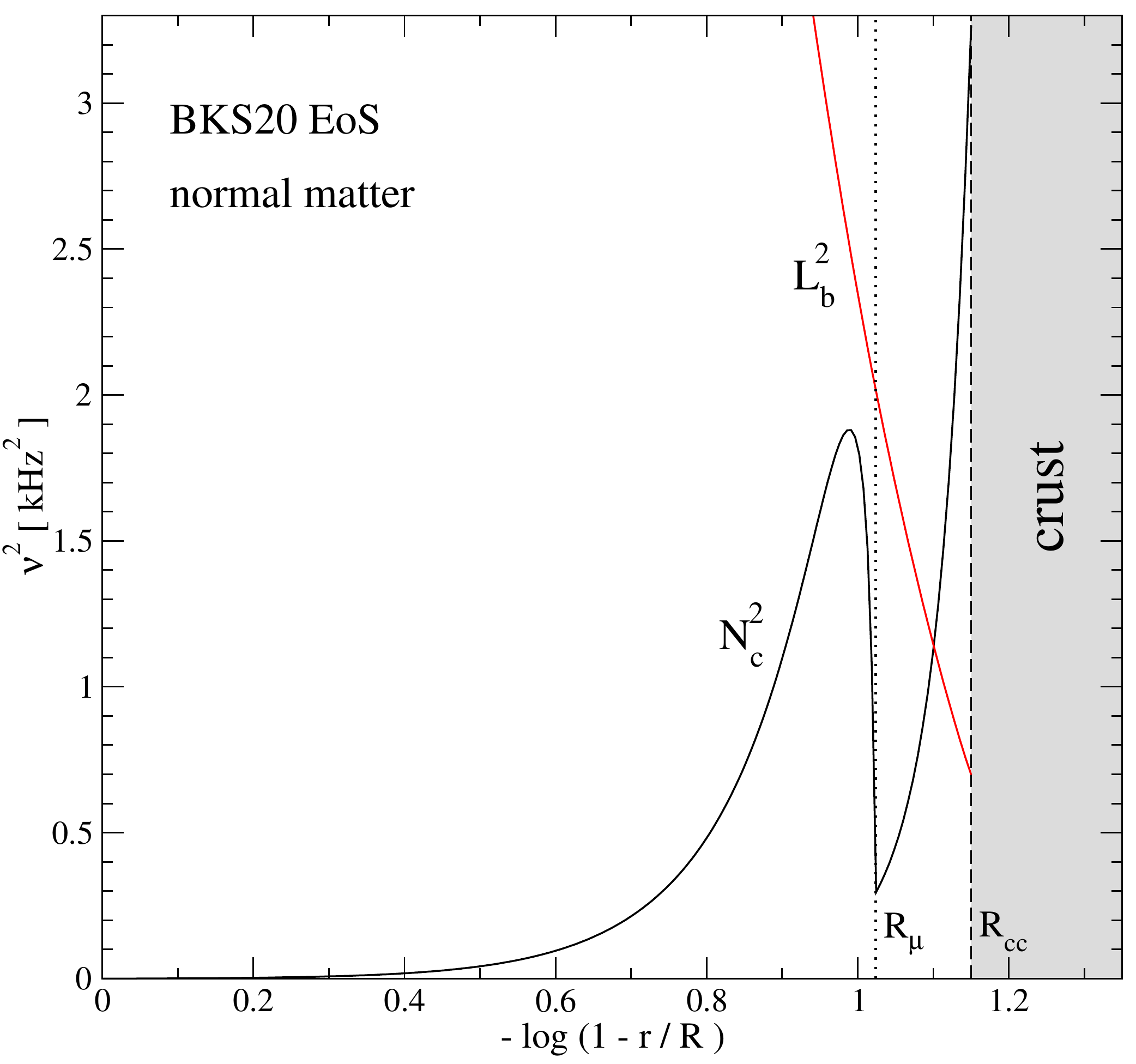} 
\caption{Propagation diagram for a cold, normal fluid, neutron star core. We show the frequencies $\nu=\sigma/2\pi$ for local sound waves (represented by $\mathcal L_b^2$ for $l=2$ obtained from \eqref{sfreq}) and the g-modes (given by $\mathcal N_c^2$ obtained from \eqref{gfreq}). The results show that the muons make the dominant contribution to the g-modes  throughout the part of the core where they are present.  
\label{fig:PD-normal}}
\end{center}
\end{figure*}
%------------------------------------------------------------------------------%

Figure~\ref{fig:PD-normal} shows numerical results for our model  star (for $l=2$). We see that the muons make the dominant contribution to the g-modes  throughout the part of the core where they are present. The upshot is that the g-mode frequency may be up to an order of magnitude higher than in pure npe matter \citep{gusk2}. We also see that (for this rather low value of $l$) the sound waves may, in fact, have lower frequency than the g-modes in the outer part of the core. This basically shows that, if we want to assume that $\mathcal L_b^2 \gg \mathcal N_c^2$, as in the above discussion, then we need to choose a larger value of $l$. It is useful to keep this in mind later.

%%%%%%%%%%%%%%%%%%%%%%%%%%%
\subsection{Thermodynamical relations and results}

Most previous g-mode calculations have been carried out using a simplified approach to the equation of state, where the stratification is accounted for by making the adiabatic index for the perturbations different from that for the background (which has often been taken to be a simple polytrope).  This approach is fine as long as one is mainly interested in qualitative properties. It is also, although perhaps to a lesser extent, natural given the difficulty to extract the required information for tabulated realistic equations of state. As we are aiming for a higher level of realism, we need to address the full problem. This means that, in order to obtain the results shown in Figure~\ref{fig:PD-normal}, we had to work out various partial derivatives for our equation of state. As this is an important step, and one has to work things out in a thermodynamically consistent way, let us consider this problem in more detail. For clarity, we focus on the region where all four particle species are present.  

First of all, we have the variation of the energy density (taken to be in the original form we assumed when we determined the background configuration)
\begin{equation}
d\veps = \sum_{\x=\n,\p,\e,\mu} \mu_\x dn_\x = {p+\veps \over n} dn + n(\mu_\p-\mu_\n) dx_\p + n\mu_\e dx_\e + n\mu_\mu dx_\mu \ ,
\label{epsvar}\end{equation}
where we have used  the (integrated) first law \eqref{firstlaw}. From this relation, it is easy to see that 
\begin{equation}
c_s^2 = \left( {\partial p \over \partial \veps}\right)_{x_\x} = {n\over w} \left( {\partial p \over \partial n}\right)_{x_\x} = {p\over w}  \Gamma_1^b \ .
\end{equation}
Hence, we see that our sound-wave solutions can be written
\begin{equation}
\sigma^2 = \mathcal L_b^2 = {l(l+1) \over r^2 }e^{2\Phi} \left[ {p\over w} \Gamma_1^b \right] = {l(l+1) \over r^2 }e^{2\Phi} \left[ c_s^2\right] \ ,
\end{equation}
which  means that our notation makes sense. It is also worth noting that the factor in the square brackets can be evaluated without reference to a specific stellar model.

This comparison was quite straightforward. However, in order to evaluate, for example, \eqref{Gammax} we need to hold the pressure $p$ fixed in the variation. This means that we need to ensure that $dp =0$.  We can recast this constraint into a relation that gives $dn$ in terms of the variations in the particle fractions;
\begin{equation}
dn = - \left( {\partial p \over \partial n } \right)_{x_\x}^{-1} \sum_{\x=\p,\e,\mu} \left( {\partial p \over \partial x_\x} \right)_{n,x_\y} 
dx_\x , \qquad \y\neq\x \ .
\end{equation}
Using this in \eqref{epsvar} we can calculate the different partial derivatives required to determine the individual $\Gamma_\x$. These can then be used in \eqref{gfreq} to obtain the (local) g-mode frequencies for a given stellar model. Adding the different terms together we get (note: the variables that are held fixed are not stated explicitly here)
\begin{multline}
\veps  \sum_{\x=\p,\e,\mu}  \frac{ x_{\x}^{\prime} }{\Gamma_{\x}}  = \veps\left[ \left( {1\over \Gamma_\p} + {1\over \Gamma_\e} \right) x_\p^\prime
+ \left( {1\over \Gamma_\mu} - {1\over \Gamma_\e} \right) x_\mu^\prime \right] 
=- {p+\veps \over n}  \left( {\partial p \over \partial n } \right)^{-1} \left[   \left( {\partial p\over \partial x_\p} + {\partial p\over \partial x_\e} \right) x_\p^\prime
+\left( {\partial p\over \partial x_\mu} - {\partial p\over \partial x_\e} \right) x_\mu^\prime   \right] \ .
\end{multline}
In the first equality we have assumed that the background is charge neutral \eqref{three}. In the second equality we have imposed the background beta-equilibrium \eqref{one} as well as the balance between muons and electrons \eqref{two}. Of course, as we have already discussed, the background composition can be expressed in terms of the baryon number density, so in fact we have 
$x_\p=x_\p(n)$, and similar for all other background quantities, which means that the final expression we need in \eqref{gfreq} is
\begin{equation}
{\veps }  \sum_{\x=\p,\e,\mu}  \frac{ x_{\x}^{\prime} }{\Gamma_{\x}}  
=- {w \over n}  \left( {\partial p \over \partial n } \right)^{-1} \left[   \left( {\partial p\over \partial x_\p} + {\partial p\over \partial x_\e} \right) {dx_\p \over dn}
+\left( {\partial p\over \partial x_\mu} - {\partial p\over \partial x_\e} \right) {dx_\mu \over dn}   \right] n^\prime \ .
\label{diffo}\end{equation}
In order to quantify the g-mode frequencies, and obtain results like those in Figure~\ref{fig:PD-normal}, we obviously need to build a stellar model. However, it is worth noting that,  if the focus is on the relative contributions of the proton and muon gradients, then all we need to do is compare the two terms in the square bracket in \eqref{diffo}. This can be done without reference to a particular star as it only requires the partial derivatives and the known composition as a function of $n$. Specifically, we first note that 
\begin{equation}
n' = \left( {dp \over dn} \right)^{-1}_\mathrm{eq} p' = - gw \left( {dp \over dn} \right)^{-1}_\mathrm{eq} \ ,
\end{equation}
which means that we have
\begin{equation}
\sigma^2 = g^2 e^{2(\Phi-\Lambda)} \left[ \veps \left( {dp \over dn} \right)^{-1}_\mathrm{eq} \sum_\x {1\over \Gamma_\x} \left( {dx_\x \over dn} \right)_\mathrm{eq}\right] \ .
\end{equation}
To quantify the terms in the square bracket we only need the pressure and composition of matter at equilibrium. We will not provide an explicit example of this comparison, because the information it would contain is already clear from Figure~\ref{fig:PD-normal}. However, it is still useful to keep this possible comparison in mind because it would provide an immediate and simple way of checking the importance of different contributions to the buoyancy for a variety of equations of state.

%%%%%%%%%%%%%%%%%%%%%%%%%%%%%%% SEC. %%%%%%%%%%%%%%%%%%%%%%%%%%%%%%%%%%%%%%%%%%%
\section{The g-modes of a superfluid neutron star core} \label{sec:eqs}
%%%%%%%%%%%%%%%%%%%%%%%%%%%%%%%%%%%%%%%%%%%%%%%%%%%%%%%%%%%%%%%%%%%%%%%%%%%%%%%%

Moving on to the issue that motivated this investigation, we consider how the problem changes when the neutrons become superfluid and we also account for finite temperature effects. As we have already discussed,  the neutral superfluid component may drift relative to the charged components, which remain locked. This adds a degree of freedom to the problem. It also means that we need to adjust our notion of composition gradients. Since the neutrons now form a separate component, that can move out of the way if it is pushed, they no longer contribute to the buoyancy that the other components experience. The upshot of this is that, if we ignore thermal effects, then there is no local support for g-modes in regions where only neutrons, protons and electrons are present. The neutrons do not play a role because they are superfluid and charge neutrality requires the electron and proton fractions to be equal. Hence, there is no buoyancy and no g-modes \citep{1995A&A...303..515L,comer}. Interestingly, this means that entropy gradients will be key to determining the nature of the g-modes in this region. This was pointed out by \citet{gusk1}, who also argued that these gradients may be such that the resulting modes are unstable. Of course, at the density where muons appear the situation changes again. Beyond this density there will again be a composition gradient as the ratio of muons to protons and electrons will vary \citep{gusk2}. This is likely to overwhelm the thermal effects and should serve to suppress any instability. Our aim in the following is to test the veracity of these expectations.
 
In a region where the neutrons are superfluid, the locked components still satisfy conservations laws of the form \eqref{constot} [replacing $n$ with $n_\p, n_\e, n_\mu$ or $s=n_\s$]. Meanwhile, we have to account for the perturbed drift velocity, $\delta v^a$, in the conservation law for the neutrons. This means that we have
 \begin{equation}
e^{-\Phi}  \partial_{t}   \delta n_{\n}  + \partial_{k} \left[  n_{\n} \left(  \delta u^{k} + \delta v^{k} \right)   \right]  
+ n_{\n} \left[ \left( \Lambda' + \Phi'   + \frac{2}{r} \right)   \left(  \delta u^{r} + \delta v^{r} \right) 
+ \cot \theta  \,  \left(  \delta u^{\theta} + \delta v^{\theta} \right) \right]  = 0 \, , 
\label{nncon}\end{equation}
Adding this to the equation for the protons we have the conservation law for baryons;
\begin{equation}
e^{-\Phi}  \partial_{t}   \delta n  + \partial_{k} \left[  n \delta u^{k} + n_{\n}\delta v^{k}    \right]  
+  \left( \Lambda' + \Phi'   + \frac{2}{r} \right)   \left( n  \delta u^{r} + n_{\n} \delta v^{r} \right) 
+ \cot \theta  \,  \left( n \delta u^{\theta} + n_{\n}  \delta v^{\theta} \right)   = 0 \, .  \\
\label{nbcon}\end{equation}

If we ignore entrainment (we will account for this effect later)  then the  (perturbed) momentum  equation for the superfluid neutrons takes the very simple form;
\begin{align}
\partial_{t}  \delta \mu^{\x}_{k}  =   \partial_{k}  \delta \mu^{\x}_{t} \, . 
\end{align}
That is, we have
\begin{equation}
\mu_{\n}  \left( \partial_{t}   \delta u_{k} + \partial_{t}   \delta v_{k} \right) = - e^{\Phi} \partial_{k} \delta \mu_{\n}  -  \delta \mu_{\n} e^{\Phi} \partial_{k} \Phi  = - \partial_k \left( e^\Phi \delta \mu_\n \right)  \, , \label{eq:M2}
% = - \partial_k \delta \tilde\mu_\n
\end{equation} 
where the last identity holds since we are working in the Cowling approximation where the metric is held fixed.  The final equation we need represents total momentum conservation. As the neutrons are allowed to drift relative to the other component this relation is different from \eqref{momtot}. We now have                                                     
\begin{equation}
w \partial_{t}   \delta u_{k} +n_{\n} \mu_\n  \partial_{t}   \delta v_{k} = - e^{\Phi}  \partial_{k} \delta p   -  e^{\Phi} \partial_{k} \Phi  \delta w  \, , \label{eq:sM1}
\end{equation} 
where we have (again) used the  enthalphy $w=p+\varepsilon$.                                                        
             
In this problem we need to keep track of two Lagrangian displacements. Following previous work on the problem, we take these to be $\xi^\alpha$ and $\eta^\alpha$,  defined such that (cf. the velocity dependence in \eqref{nncon} and \eqref{nbcon});
\begin{equation}
 e^{-\Phi} \partial_{t} \xi^{a }     = \delta u^{a} +  x_{\n} \delta v^{a} \, ,
\label{x1} \end{equation}
 and
 \begin{equation}
e^{-\Phi}  \partial_{t} \eta^{a }  = \delta u^{a} +  \delta v^{a} \ . 
\label{x2}\end{equation}
This means that (\ref{eq:M2})-(\ref{eq:sM1}) become, respectively
\begin{align}
& \alpha_{\ch}  \frac{ \partial ^2  \xi_{k} }{ \partial t ^2} - \alpha_{\n} \frac{ \partial ^2  \eta_{k} }{ \partial t ^2}  = - e^{2 \Phi}  \partial_{k} \delta p   -   g e^{2 \Phi}  \delta w  \label{eq:sM1a} \,  ,  \\ 
& \frac{ \partial ^2  \eta_{k} }{ \partial t ^2} =  - e^{2 \Phi}  \partial_{k} \delta \mu_{\n}^{\star} \, ,  \label{eq:sM2a}
\end{align} 
 where we have defined
 \begin{equation}
\delta \mu_{\n}^{\star} = \frac{ \delta \mu_{\n} }{\mu_{\n}} \, .
\end{equation} 
We have also used the background equation (\ref{eq:back}). Finally, we have  defined
\begin{equation}
 \alpha_{\ch}  =   \frac{ w - n_{\n} \mu_{\n} }{x_{\p}} \approx w(T=0)+{sT\over x_\p}  \, , 
\end{equation}
and
\begin{equation}
\alpha_{\n}    =   \frac{ x_{\n} w - n_{\n} \mu_{\n}}{x_{\p}} = \frac{x_\n}{x_\p} \left( w - n \mu_\n\right)\approx {x_\n \over x_\p} sT \ . 
\end{equation}
The last equalities assume the cold equilibrium conditions for the chemical potentials, which is somewhat inconsistent as we would then have to have $T=0$. However, the purpose here is to argue that $\alpha_\n\ll\alpha_\ch$ which means that we can often neglect this contribution. Having said that, we will retain this quantity throughout the calculation because this will  facilitate a comparison with the case where the entrainment is accounted for later.

\subsection{The dispersion relation (no entrainment)}

In order to derive the dispersion relation for the superfluid problem, we (again) assume time dependence $e^{i\sigma t}$ and decompose the variables in spherical harmonics. The displacement $\xi^a$ is decomposed as in \eqref{Wdef}-\eqref{Vdef2}, while $\eta^a$ takes the same form but with $W_\n$ and $V_\n$ replacing $W$ and $V$. As in the analysis of the normal-fluid problem in Section~3, we  use hats to denote the amplitudes of various perturbed scalar quantities.

The two scalar conservation laws then lead to 
\begin{align}
& W^{\prime}+ \left[ \frac{n^{\prime}}{n}  + \frac{2}{r} + {\Lambda'} \right] W =  \frac{l(l+1) }{ r^2} V -   \frac{ \hat n }{n} \label{eq:sM3} \, , \\
& W^{\prime}_{\n}+ \left[ \frac{n^{\prime}_{\n}}{n_{\n}}  + \frac{2}{r} + {\Lambda'} \right] W_{\n} =  \frac{l(l+1) }{ r^2} V_{\n} -   \frac{ \hat n_{\n} }{n_{\n}} \label{eq:sM4} \, ,
\end{align}
while the radial components of the momentum equations become;
\begin{align}
&  \sigma^2  \left( \alpha_{\ch}   W  - \alpha_{\n}    W_{\n}  \right) e^{-2(\Phi-\Lambda)}  =  \partial_{r} \hat p   +   g  \hat w  \,  ,  \label{eq:sM1b} \\
&  \sigma^2   W_{\n}   e^{-2(\Phi-\Lambda)}  =   \partial_{r} \hat \mu_{\n}^{\star} \, , \label{eq:M2b} 
\end{align} 
and the angular parts give;
\begin{align}
&  \sigma^2  \left( \alpha_{\ch}   V  - \alpha_{\n}    V_{\n}  \right) e^{-2 \Phi}  =  \hat p   \,  ,  \\
&  \sigma^2   V_{\n}   e^{-2 \Phi}  =  \hat \mu_{\n}^{\star} \, .
\end{align} 
Combining these we have
\begin{equation}
  V     =    \frac{ \hat p  + \alpha_{\n} \hat \mu_{\n}^{\star} }{   \alpha_{\ch} \sigma^2   }  \, e^{ 2 \Phi} \ , 
\end{equation}
and we see that we can remove both $V$ and $V_\n$ from the analysis at this point.

Let us now assume that we focus on the part of the star's core where muons are present, and that we choose to work with $p$, $\mu_\n$ and $S=s/n_\e$ and $M=n_\mu/n_\e$ as our primary variables\footnote{Note that $S$ is defined as the entropy per electron. This makes sense because the electrons and muons carry the entropy when the neutrons are superfluid and the protons are superconducting. We also need the ratio to be defined between two co-moving components in order to show that the Lagrangian variation of this quantity vanishes, cf. Section~3.1.}. These are the same variables as \citet{gusk1} used, which facilitates a direct comparison. We assume that the perturbations are adiabatic, which means that 
\begin{equation}
\Delta S = 0 \quad \longrightarrow \quad \delta S = - \xi^{r}_{S} S'  = \frac{n_\n W_{\n} - n  W }{ n_\ch }  S^{\prime} \ ,
\label{entper}
\end{equation}
where  $\xi^r_S$ is the radial component of the displacement associated with $\delta u^a$ (and can be obtained by inverting \eqref{x1} and \eqref{x2}). If the muons and electrons are also locked together (as we assume), then the relation for $\delta M$ is obtained simply by replacing $S$ with $M$ in the above result. In fact, as these two quantities enter the equations that follow in a similar way it is convenient to introduce a label $X$ which can be either $S$ or $M$ (and also use $Y\neq X$). 

Making use of thermodynamics, we have for the enthalpy:
\begin{equation}
\delta w = \delta \veps + \delta p = \left( 1 + {1\over \bar c_s^2} \right) \delta p + {\veps\over \bar \Gamma_\mu} \delta \mu_\n^{\star} - \frac{\veps }{ g} N_c^2 \left( x_{\n} W_{\n} - W \right) \, , 
\label{enthvar}\end{equation}
where the term that ultimately leads to the buoyancy takes the form
\begin{equation}
N_c^2  = - \frac{g}{x_{\p} } \sum_X  \frac{X^{\prime} }{\bar \Gamma_{X}}  \, .
\end{equation}
Here, and in the following, we use bars over the adiabatic indices (eg. $\bar \Gamma_X$) to make a distinction from the analogous quantities in the normal-fluid case discussed earlier. 
It is important to make this clear as we use different variables and different quantities are held fixed in the partial derivatives. 

As in the case of a system where all components are locked together, the buoyancy depends on a balance of terms. In the present case we have thermal stratification and the muon gradient. It is worth noting that overall charge neutrality implies that 
\begin{equation}
M^\prime = \left( {n_\mu \over n_\e} \right)^\prime =  \left( {n_\p \over n_\e} -1 \right)^\prime \ ,
\end{equation}
from which it is easy to see that the second contribution to $N_c^2$ vanishes in the region where the muons are absent (as it must). 

In \eqref{enthvar} we have used
\begin{equation}
\bar c_s^2 = \left( {\partial p \over \partial \veps} \right)_{\mu_\n, X} \ ,
\end{equation}
(as before, the bar is used to make a distinction between the superfluid and the normal-fluid problem)
\begin{equation}
{1\over \bar \Gamma_\mu} = {\mu_\n \over \veps}  \left( {\partial \veps \over \partial \mu_\n} \right)_{p, X} \ ,
\end{equation}
(where the index $\mu$ on $\Gamma$ refers to the chemical potential rather than the muons)
and
\begin{equation}
{1\over \bar \Gamma_X} = {1 \over \veps}  \left( {\partial \veps \over \partial X} \right)_{p, \mu_\n, Y} \ .
\end{equation}

Similarly, the  neutron number density perturbation is now given by
\begin{equation}
 \frac{ \delta n_\n }{n_\n}  =   \frac{1}{\bar  \Gamma_1^\n } \frac{\delta p}{p} +  \frac{1}{ \bar \Gamma_\mu^\n } \delta \mu_\n^{\star} +   \frac{n_\n W_{\n} - n  W }{ n_\p } \mathcal A_\n  \, , 
\end{equation}
with 
\begin{equation}
{1\over \bar \Gamma_1^\n} = {p \over n_\n}  \left( {\partial n_\n \over \partial p} \right)_{\mu_\n, X} \ ,
\end{equation}
(where the $\n$ index indicates that these adiabatic indices are associated with the variation of the neutron number density $n_\n$)
\begin{equation}
{1\over \bar \Gamma_\mu^\n} = {\mu_\n \over n_\n}  \left( {\partial n_\n \over \partial \mu_\n} \right)_{p, X}  \ ,
\end{equation}
and, for later convenience, we have defined
\begin{equation}
\mathcal A_\n =  \sum_X \frac{ {X}^{\prime} }{ \bar \Gamma_{X}^\n}   \, .
\end{equation}
with
\begin{equation}
{1\over \bar \Gamma_X^\n} ={1\over n_\n}   \left( {\partial n_\n \over \partial X} \right)_{p,\mu_\n, Y}  \ .
\end{equation}

Finally, we have the baryon number perturbation;
\begin{equation}
 \frac{ \delta n }{n }  
=   \frac{1}{\bar  \Gamma_1^b } \frac{\delta p}{p} +  \frac{1}{ \bar \Gamma_\mu^b } \delta \mu_\n^{\star} +   \frac{n_\n W_{\n} - n  W }{ n_\p } \mathcal A_{b}  \, , 
 \end{equation}
 where we have used
 \begin{equation}
{1\over \bar \Gamma_1^b} = {p \over n}  \left( {\partial n \over \partial p} \right)_{\mu_\n, X}   \ ,
\end{equation}
 \begin{equation}
{1\over \bar \Gamma_\mu^b} = {\mu_\n \over n}  \left( {\partial n \over \partial \mu_\n} \right)_{p, X} \ ,
\end{equation}
and
\begin{equation}
 \mathcal A_b =  \sum_X \frac{ X^{\prime} }{\bar\Gamma_X^b}  \, ,
\end{equation}
with
\begin{equation}
{1\over \bar \Gamma_X^b} ={1\over n}   \left( {\partial n \over \partial X} \right)_{p,\mu_\n,Y} \ .
\end{equation}

As in the normal-fluid case, it is natural to introduce;
\begin{equation}
\mathcal L_b^2 =  \frac{p}{\alpha_{\ch}}   \frac{l(l+1) }{ r^2 } e^{2\Phi}  \bar \Gamma_1^b   \ .
\label{Lb2}\end{equation}
Note that, all quantities labelled $\mathcal L$ (with various indices) contain the ``centrifugal'' $l(l+1)$ factor. They do not, in the general setting, directly correspond to solutions to the final dispersion relation. That was particular to the normal-fluid problem.

Let us now manipulate  equation (\ref{eq:sM3});
\begin{equation}
 W^{\prime}+ \left[ \frac{n^{\prime}}{n}  + \frac{2}{r} + {\Lambda'} - \frac{\mathcal A_b   }{x_{\p}}\right] W =  
\frac{  \mathcal L_b^2 - \sigma^2 }{\bar\Gamma_1^b \sigma^2 } \frac{\hat p }{p}
+ \left[  \frac{l(l+1) }{ r^2}   \frac{ \alpha_{\n} e^{ 2 \Phi}}{   \alpha_{\ch} \sigma^2   }   - \frac{1}{ \bar \Gamma_\mu^b } \right]   \hat \mu_{\n}^{\star}    
-   \frac{n_\n W_{\n}  }{ n_\p } \mathcal A_b  \ ,
\label{Weq}
\end{equation}
Similarly, equation (\ref{eq:sM4}) becomes
\begin{equation}
W^{\prime}_{\n}+ \left[ \frac{n^{\prime}_{\n}}{n_{\n}}  + \frac{2}{r} + {\Lambda'}  +  \frac{n_\n }{ n_\p } \mathcal A_\n \right] W_{\n} = \frac{\mathcal{L}_{\n}^2 - \sigma^2 }{\sigma^2 \bar \Gamma^\n_\mu }   \hat \mu_\n^{\star}  
 -   \frac{1}{ \bar \Gamma_1^\n } \frac{\hat p}{p}   +   \frac{  \mathcal A_\n}{ x_\p } W \, ,
 \label{Wneq}   
 \end{equation}
where we have defined
\begin{equation}
\mathcal L_\n^2 =  \frac{l(l+1) }{r^2 } e^{ 2 \Phi} \bar \Gamma^\n_\mu  \, .
\end{equation}

Next  consider the radial part of (\ref{eq:sM1b});
\begin{equation}
\partial_{r} \hat p   +   g  \left( 1 + {1\over \bar c_s^2} \right) \hat p  =   \alpha_{\ch}  e^{-2(\Phi-\Lambda)}  \left[ \left( \sigma^2 -  \mathcal N_c^2 \right)  W    
+      \left(  x_{\n} \mathcal{N}_c^2 -   \sigma^2  \frac{ \alpha_{\n} }{\alpha_{\ch}}         \right) W_{\n} \right]  - g  {\veps\over \bar \Gamma_\mu} \hat \mu_\n^{\star} \,  ,  
\label{peq}\end{equation} 
where we have defined (cf. the analogous definition in the cold, normal fluid problem, Eq.~\eqref{normbrunt})
\begin{equation}
\mathcal N_c^2 =  \frac{\veps}{\alpha_{\ch}}  e^{2(\Phi-\Lambda)}  N_c^2 \, ,
\end{equation} 
Finally, the  momentum equation for the neutrons is simple:
\begin{equation}
 \partial_{r} \hat \mu_{\n}^{\star}  =   \sigma^2   W_{\n}   e^{-2(\Phi-\Lambda)}  \, , 
\label{mneq}\end{equation} 

The four equations \eqref{Weq}, \eqref{Wneq}, \eqref{peq} and \eqref{mneq} provide us with all the information we need to determine the dispersion relation in this more complex setting. As in the single-fluid problem, we can introduce integrating factors to simplify the equations\footnote{The integrating factors for equations (\ref{Weq}), (\ref{Wneq}) and (\ref{peq}) are, respectively, given by
\begin{equation}
 n \,  r^2  \exp \left[  \Lambda  - \int  \frac{\mathcal{A}_b }{x_\ch}   dr  \right]   \, , \quad 
n_\n \,  r^2  \exp \left[  \Lambda  - \int  \frac{n_\n}{n_\ch}  \mathcal{A}_\n  dr  \right]   \, ,  \quad \mbox{and} \quad
  \exp \left[  \int   g \left( 1 + \frac{1}{\bar c_s^2} \right) dr \right]\ .
\end{equation} } and yet again the upshot of this is that we can effectively ignore the terms in brackets on the left-hand sides of \eqref{Weq}, \eqref{Wneq} and \eqref{peq}. Making this simplification and focussing on plane-wave solutions (i.e. taking the perturbations to depend on the radial coordinate as  $ ~ e^{i k r}$, but not changing the notation for the amplitudes, as there should be no risk of confusion) we have

\begin{align}
&  i k \, W   =  
\frac{  \mathcal L_b^2 - \sigma^2 }{\bar\Gamma_1^b \sigma^2 } \frac{\hat p }{p}
+ \left[  \frac{l(l+1) }{ r^2}   \frac{ \alpha_{\n} e^{ 2 \Phi}}{   \alpha_{\ch} \sigma^2   }   - \frac{1}{ \bar  \Gamma_\mu^b } \right]   \hat \mu_{\n}^{\star}    
  -   \frac{n_\n \mathcal A_b    }{ n_\p } W_{\n}  \, , \\
&  i k \, W_{\n}  = \frac{\mathcal{L}_{\n}^2 - \sigma^2 }{\sigma^2 \bar \Gamma^\n_\mu}   \hat \mu_\n^{\star}  
 -   \frac{1}{ \bar \Gamma_1^\n } \frac{\hat p}{p}   +   \frac{  \mathcal A_\n}{ x_\p } W    \, , \\
&  i k  \, \hat p     =   \alpha_{\ch}  e^{-2(\Phi-\Lambda)}  \left[ \left( \sigma^2 -  \mathcal N_c^2 \right)  W    
+      \left(  x_{\n} \mathcal{N}_c^2 -   \sigma^2  \frac{ \alpha_{\n} }{\alpha_{\ch}}         \right) W_{\n} \right]  -  \frac{g \veps}{ \bar \Gamma_\mu} \hat \mu_\n^{\star} \,  ,  \\
 &  i k  \, \hat \mu_{\n}^{\star}  =   \sigma^2   W_{\n}   e^{-2(\Phi-\Lambda)}  \,  .   \label{eq:Wn}
\end{align} 

The steps required to obtain the dispersion relation are straightforward. First we remove $W$ and $W_\n$ from the equations to get two coupled equations for $\hat p$ and $\hat \mu_\n^\star$. Finally, combining those equations, we arrive at the dispersion relation
\begin{multline}
 \left[ \tilde k^2 + \frac{ \alpha_\ch}{p} \left( \sigma^2 -  \mathcal N_c^2 \right)   \frac{  \mathcal L_b^2 - \sigma^2 }{\bar\Gamma_1^b \sigma^2 }  \right] \times
  \left(  \tilde k^2   + \frac{\mathcal{L}_{\n}^2 - \sigma^2 }{\bar \Gamma^\n_\mu}   \right)    \\
  =  \left[ {\mathcal A_\n \tilde k^2 \over x_\p} + {ik\over p \bar  \Gamma_1^\n} \alpha_\ch \left( \sigma^2 - \mathcal N_\ch^2\right) \right]\times
  \left[ \left(   \mathcal{L}_{\star}^2 - \sigma^2   \right) \frac{i}{ k  \bar  \Gamma_{\mu}^b } 
+  \frac{x_\n}{x_\p} \mathcal A _b  e^{2(\Phi-\Lambda)}  \right] \\
- {i\tilde k^2 \over k} \left[  \alpha_\ch x_\n \mathcal N_\ch ^2 - \alpha_\n\sigma^2 + \frac{i\sigma^2}{k}   \frac{g \veps}{ \bar  \Gamma_\mu} \right]   \left(  \frac{i k}{ p \bar  \Gamma_1^\n }       
-  \frac{  \mathcal A_\n}{x_\p p} \frac{  \mathcal L_b^2 - \sigma^2 }{\bar \Gamma_1^b \sigma^2 }  \right)    \ ,
\label{bigdisp}\end{multline}
where we have introduced 
\begin{equation}
 \mathcal{L}_{\star}^2 =   \bar  \Gamma_\mu^b \frac{l \left( l + 1 \right)}{r^2} \frac{\alpha_\n}{\alpha_\ch} e^{2 \Phi}  \ ,
\end{equation}
and  used $\tilde k = k e^{\Phi - \Lambda}$, as before.
It is worth noting that the dispersion relation is now a cubic in $\sigma^2$, which means that we  expect to find three (more or less distinct) classes of waves. In addition to the acoustic modes and the g-modes from the single-fluid problem we should have a set of superfluid modes \citep{comer}. 

%------------------------------FIG. 4------------------------------------------%
\begin{figure*}
\begin{center}
\includegraphics[height=75mm]{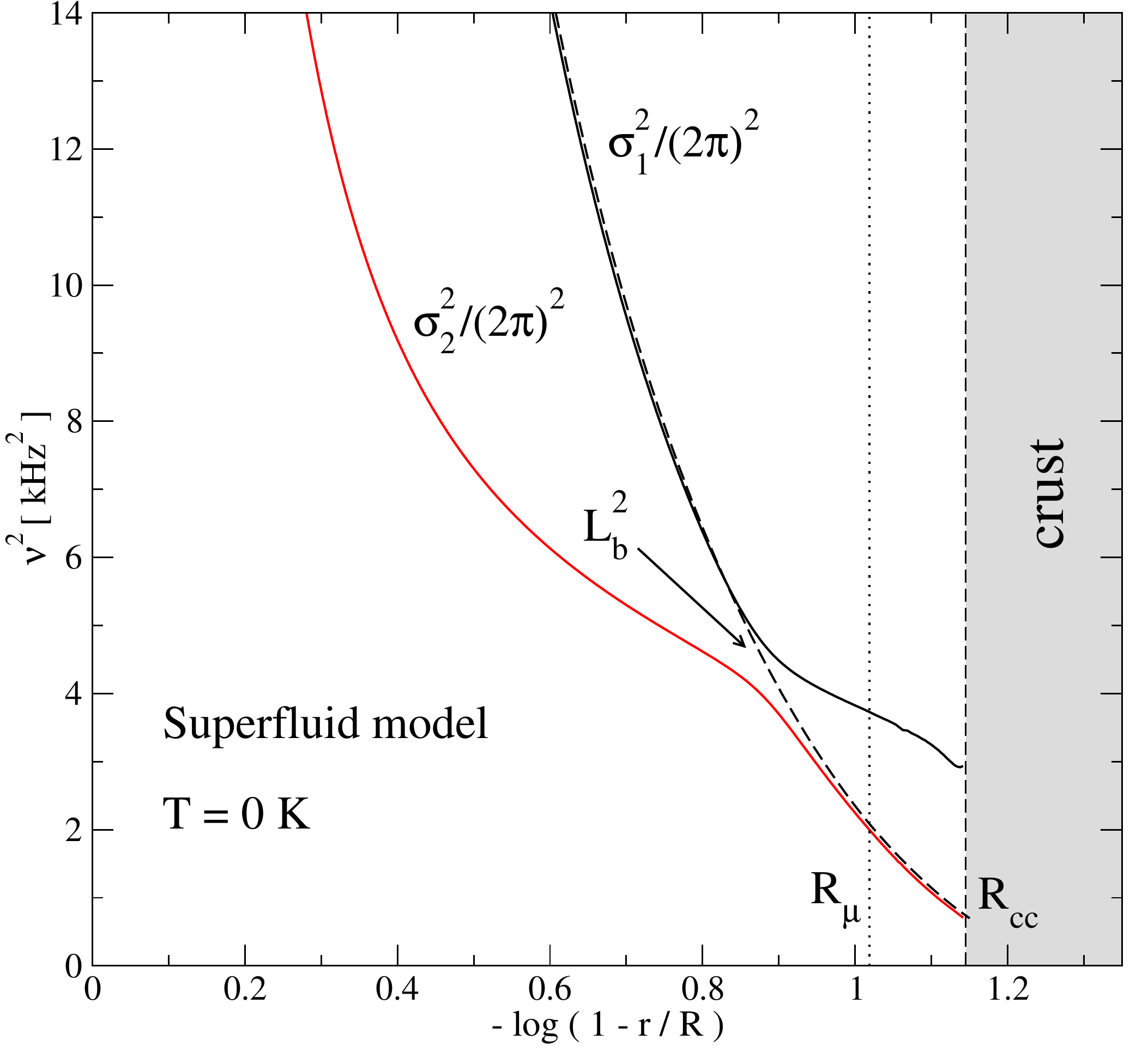}
\caption{ The two sound-wave solutions that follow from the simplified dispersion relation (\ref{eq:DR2}) for the superfluid problem (and $l=2$), assuming that the star is sufficiently cold that the entire core is superfluid. Note that the solutions show an avoided crossing in the outer core of the star. We also show (as a dashed line) the speed of sound from the normal-fluid problem ($L_b^2$ from \eqref{sfreq}). The results show that one of the two sound speeds in the superfluid problem tends to be close to the ``normal sound'', but a switch-over takes place at the avoided crossing.
\label{sfsound}}
\end{center}
\end{figure*}
%------------------------------------------------------------------------------%

\subsection{Results for a cold superfluid core}

The result \eqref{bigdisp}  provides the complete dispersion relation, without approximations other than the plane-wave assumption. Let us now add to this the assumption that we are dealing with short-wavelength/large multipole waves, such that $1 \ll {kr} \ll l$ (as before). After comparing the different part of  \eqref{bigdisp}  and keeping only the dominant terms, we arrive at the simplified relation for high-frequency waves 
\begin{equation}
\left(  \mathcal L_b^2 -\sigma^2 \right) \left(  \mathcal L_n^2 -\sigma^2 \right)   -  \chi  \sigma^4  \simeq 0 \, .
 \label{eq:DR2}
\end{equation}
where 
\begin{equation}
\chi \equiv  \frac{  \bar \Gamma_1^b }{ \bar \Gamma_1^{\n} }  \frac{   \bar \Gamma^n_{\mu}    }{  \bar \Gamma_{\mu} ^b}  \ . 
\end{equation}
The sound waves follow from this quadratic in $\sigma^2$. Formally, we get
\begin{equation}
\sigma_{1,2}^2 = { (1\pm \psi) \mathcal L_b^2 + (1\mp \psi) \mathcal L_\n^2 \over 2(1-\chi)} \ ,
\label{twosounds}
\end{equation}
where
\begin{equation}
\psi = \left[ 1 + {4\chi \mathcal L_b^2 \mathcal L_\n^2 \over (\mathcal L_b^2 - \mathcal L_\n^2)} \right]^{1/2} \ .
\end{equation}
This result shows that the sound waves are typically some combination of $\mathcal L_b^2$ and $\mathcal L_\n^2$. Only in the particular case when $\chi=0$ do these quantities themselves represent solutions to the problem. This agrees with the results of \citet{comer}.
In practice, we also find that one has to be a little bit careful when evaluating these expression. For our chosen equation of state, the quantity $\bar \Gamma^\n_\mu$ changes sign in the outer core and this could cause numerical difficulties. The actual  sound-wave solutions are, however, regular. Results for our model star (and $l=2$) are shown in Figure~\ref{sfsound}. The frequencies generally decrease outwards in the star, as expected, and we also see an avoided crossing; a common feature of this kind of problem. Comparing to the normal fluid case, we find that one of the two solutions in the superfluid problem tends to be close to the ``usually assumed'' sound speed. However, the assignation changes at the avoid crossing, which is an interesting observation. 
 
Turning to the g-modes, the low-frequency solutions to \eqref{bigdisp} are approximately given by
\begin{equation}
\sigma^2 = \mathcal N_\ch^2 =- \frac{g \veps}{\alpha_{\ch} x_\p}  e^{2(\Phi-\Lambda)}  \left(  \frac{ S^{\prime} }{\bar \Gamma_{S}} + \frac{ M^{\prime} }{\bar \Gamma_{M}}    \right) 
=\sigma^2_S + \sigma^2_M\ ,
\label{gsuper} 
\end{equation}
where the coefficients ${\bar \Gamma_{S}} $ and ${\bar \Gamma_{M}}$ are discussed in Appendix~A. In this expression, 
 the muon contribution to the buoyancy can be written (making use of $\alpha_c=w$)
\begin{equation}
\sigma^2_M= g^2 e^{2(\Phi-\Lambda)}  \left[ {1\over x_\p} {\veps \over \bar \Gamma_M} \left({dp\over dn}\right)_\mathrm{eq}^{-1} \left({dM \over dn}\right)_\mathrm{eq}\right] \ ,
\end{equation}
while the entropy contribution takes the form
\begin{equation}
\sigma^2_S = g^2 e^{2(\Phi-\Lambda)} \frac{w}{\alpha_c} T^2 \left( {A \over x_\p} \right) \left\{ {1\over 3 n_\e} \left({dp\over dn}\right)_\mathrm{eq}^{-1}\left[   \left( {2M^{-1/3}\over 1+M^{2/3}}\right) \left({dn_\mu\over dn}\right)_\mathrm{eq}   - \left( { 1 + 3M^{2/3}\over 1+M^{2/3}}\right) \left({dn_\e\over dn}\right)_\mathrm{eq}  \right] - {T'\over gw T} \right\} \ ,
\label{entro}
\end{equation}
where we have used
\begin{equation}
{\veps S \over \bar \Gamma_S} = A T^2 \ ,
\end{equation}
and the coefficient $A$ is obtained by combining \eqref{xs1} with the results from Appendix~A2. 

%------------------------------FIG. 5------------------------------------------%
\begin{figure*}
\begin{center}
\includegraphics[height=70mm]{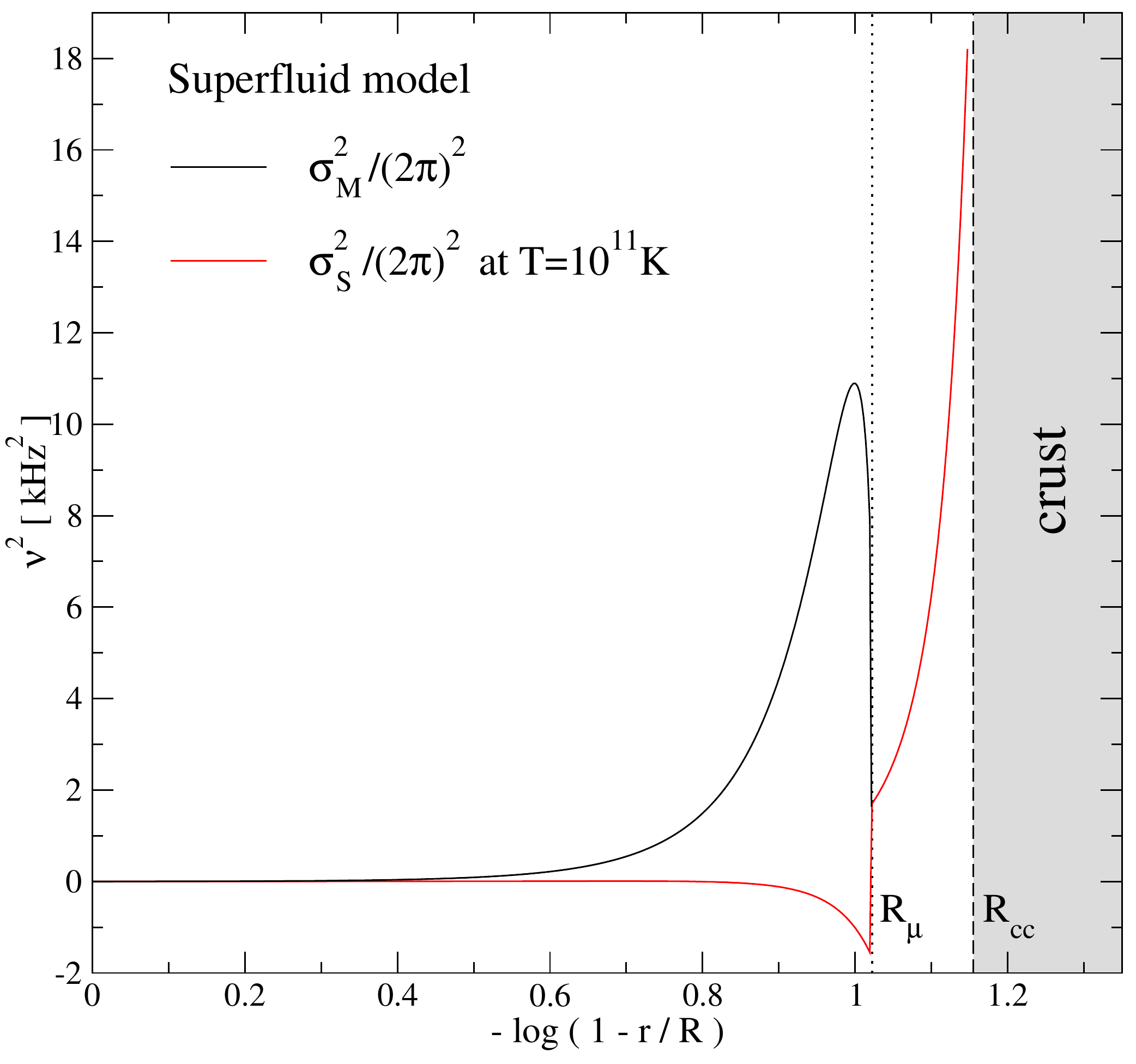}  
\caption{A comparison of the two contributions to the superfluid g-modes. The results show that the  muons contribution $\sigma^2_M$ always dominates over the thermal part $\sigma^2_S$, even if we  artificially inflate the temperature to $10^{11}$~K, an order of magnitude above our initial value for the cooling simulations. \label{compo}}
\end{center}
\end{figure*}
%------------------------------------------------------------------------------%

\begin{figure*}
\begin{center}
\includegraphics[height=70mm]{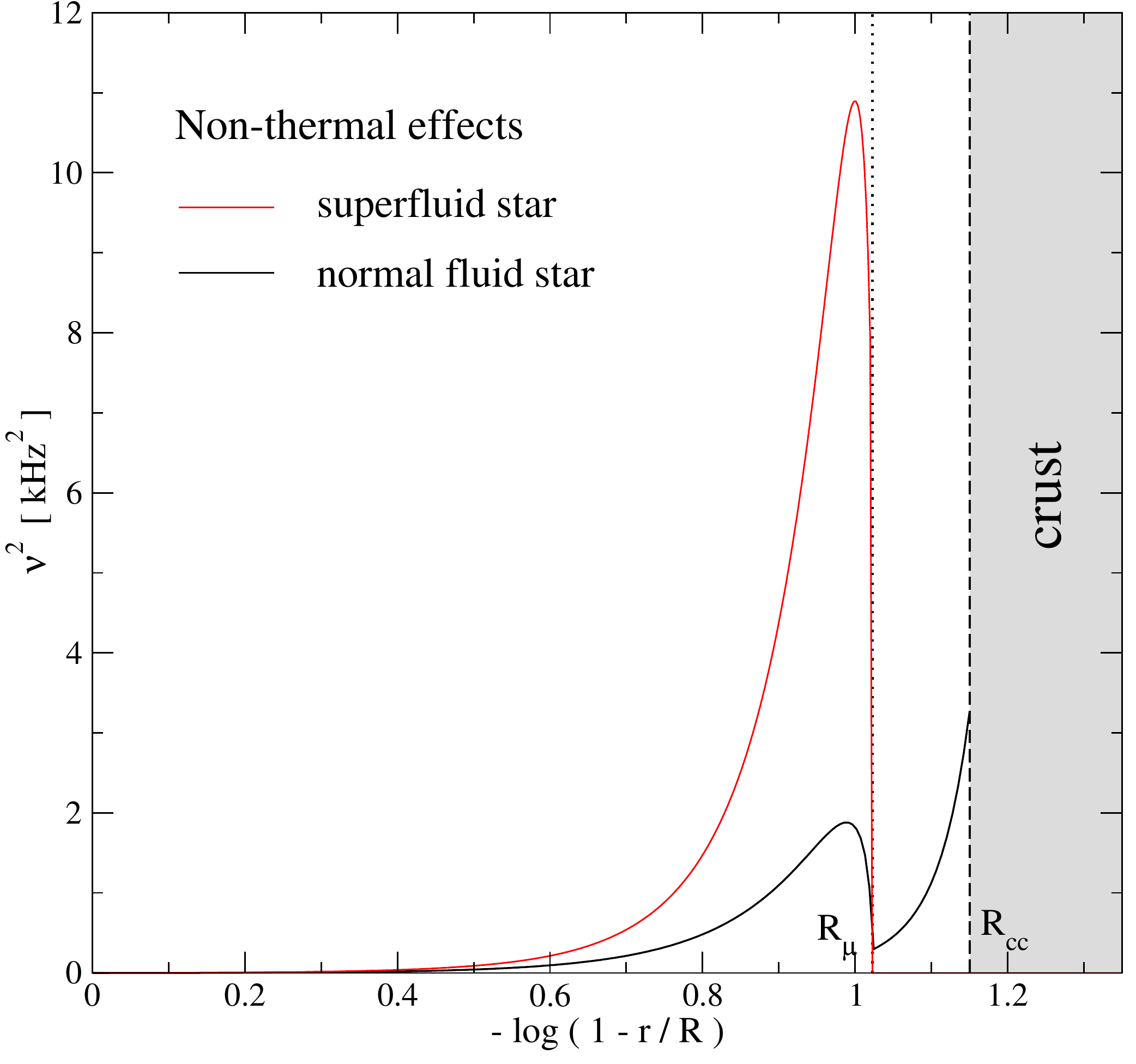}  
\caption{A comparison of the superfluid g-modes (from \eqref{gsuper}) and the normal fluid results (from \eqref{gfreq}), assuming that the star is cold so that thermal effects can be ignored. In the superfluid case there is no support for g-modes until the muons appear. In the region where the muons are present, the g-mode frequency is found to be significantly higher in the superfluid case. This is simply understood from the fact that the neutrons have been decoupled, which means that the mass involved in the wave differs by a factor of (roughly) $x_\p$.
\label{comp2}}
\end{center}
\end{figure*}

Making use of these results, we can carry out two comparisons. First, we can establish that the muons contribution $\sigma^2_M$ always dominates over the thermal part $\sigma^2_S$. This is shown in Figure~\ref{compo} where we have artificially inflated the temperature to $10^{11}$~K, an order of magnitude above our initial value for the cooling simulations (in this comparison the temperature is assumed to be uniform, so we only include the first term in \eqref{entro}). Of course, the thermal contribution has no competition in the part of the region where the muons are not present, so we still need to consider it. The second comparison is between the superfluid g-modes and the normal fluid results from Section~3. This comparison is made in Figure~\ref{comp2}, assuming that the star is cold. This means that there is no support for g-modes until the muons appear. In the region where the muons are present, the g-mode frequency is found to be significantly higher in the superfluid case. This is simply understood from the fact that the neutron have been decoupled, which means that the mass involved in the wave differs by a factor of (roughly) $x_\p$. That this is, indeed, the difference between the two cases can be shown by taking a closer look at the two expressions for $\mathcal N_c^2$.

\begin{figure*}
\begin{center}
\includegraphics[height=50mm]{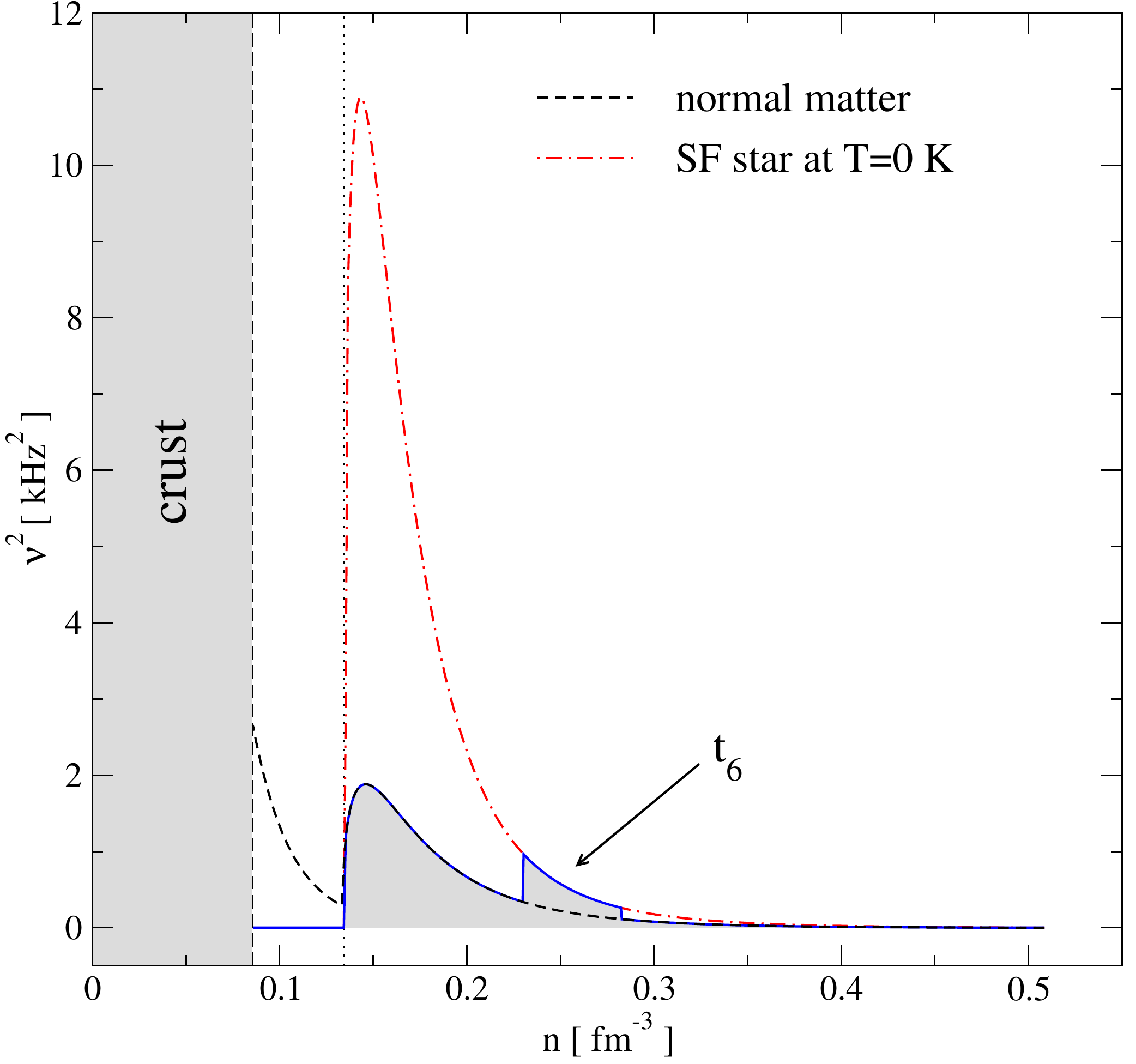}  
\includegraphics[height=50mm]{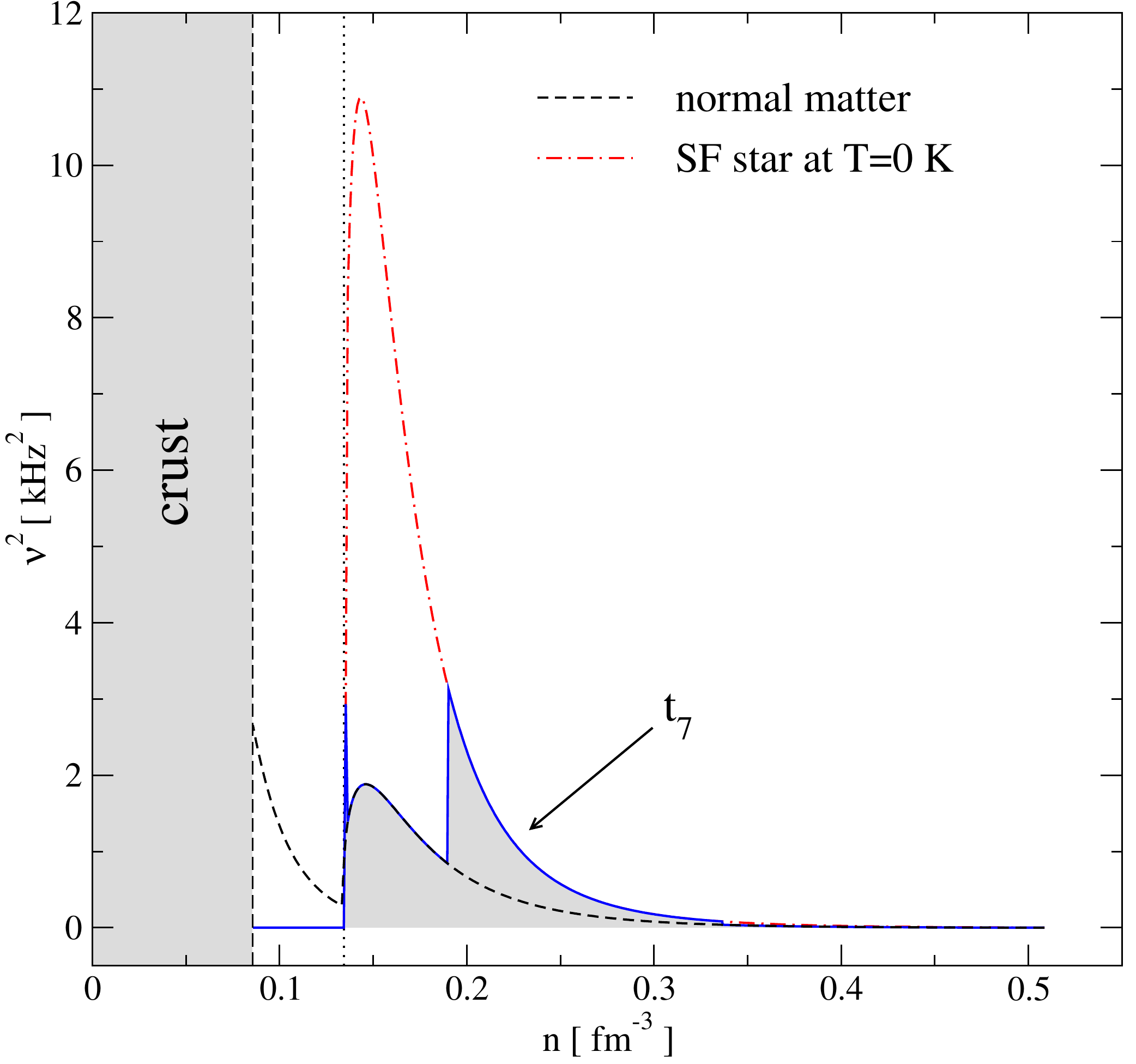}  
\includegraphics[height=50mm]{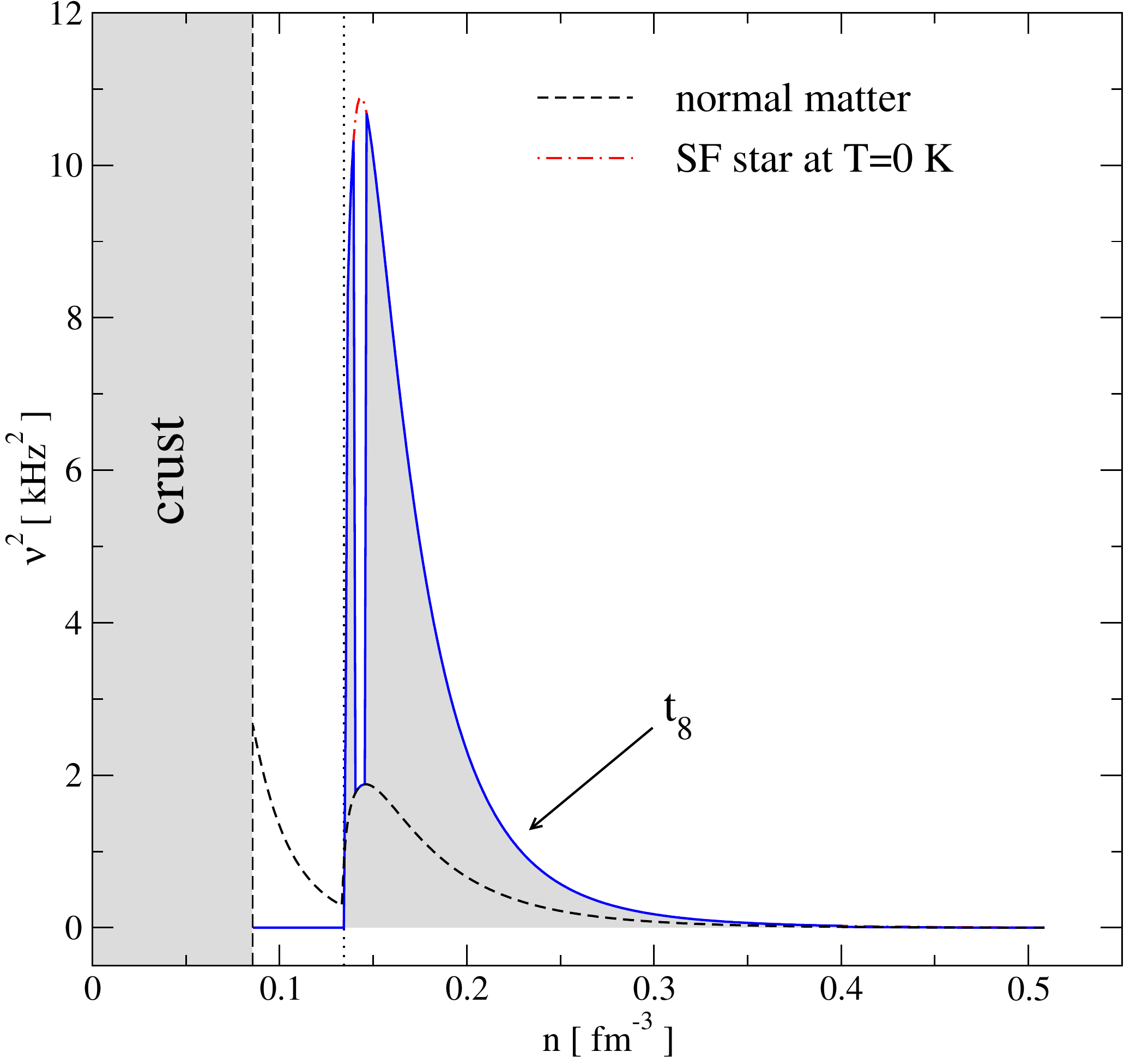}  

\caption{A sequence of snapshots corresponding to temperature distributions extracted at three particular times from Figure~\ref{fig:cooling}.  The results show how the superfluid region grows as the star cools, and how we have to replace the normal-fluid g-mode frequency from Section~3 with the superfluid results from Section~4.  The specific cooling data relates to $t_6=50$~yr, $t_7=60$~yr and $t_8=70$~yr, respectively. The corresponding temperature profiles are shown in Figure~\ref{fig:cooling}.
\label{sequence}}
\end{center}
\end{figure*}

Let us now combine our results with the cooling data from Section~2.4. That is, let us trace how the g-modes evolve as the star cools and the core becomes superfluid. If we ignore the entrainment (including thermal excitations) then the results of this exercise should be easy to understand. In a region where the fluid is above the superfluid transition temperature, the normal-fluid results from Section~3 should apply, and in superfluid regions we have to use the results from this Section. How this works out in practice is shown in Figure~\ref{sequence}. The results clearly show how the superfluid region grows as the star cools, in accordance with the results from Figure~\ref{fig:cooling}.

%------------------------------FIG. 6------------------------------------------%
%------------------------------------------------------------------------------%

\begin{figure*}
\begin{center}
\includegraphics[height=75mm]{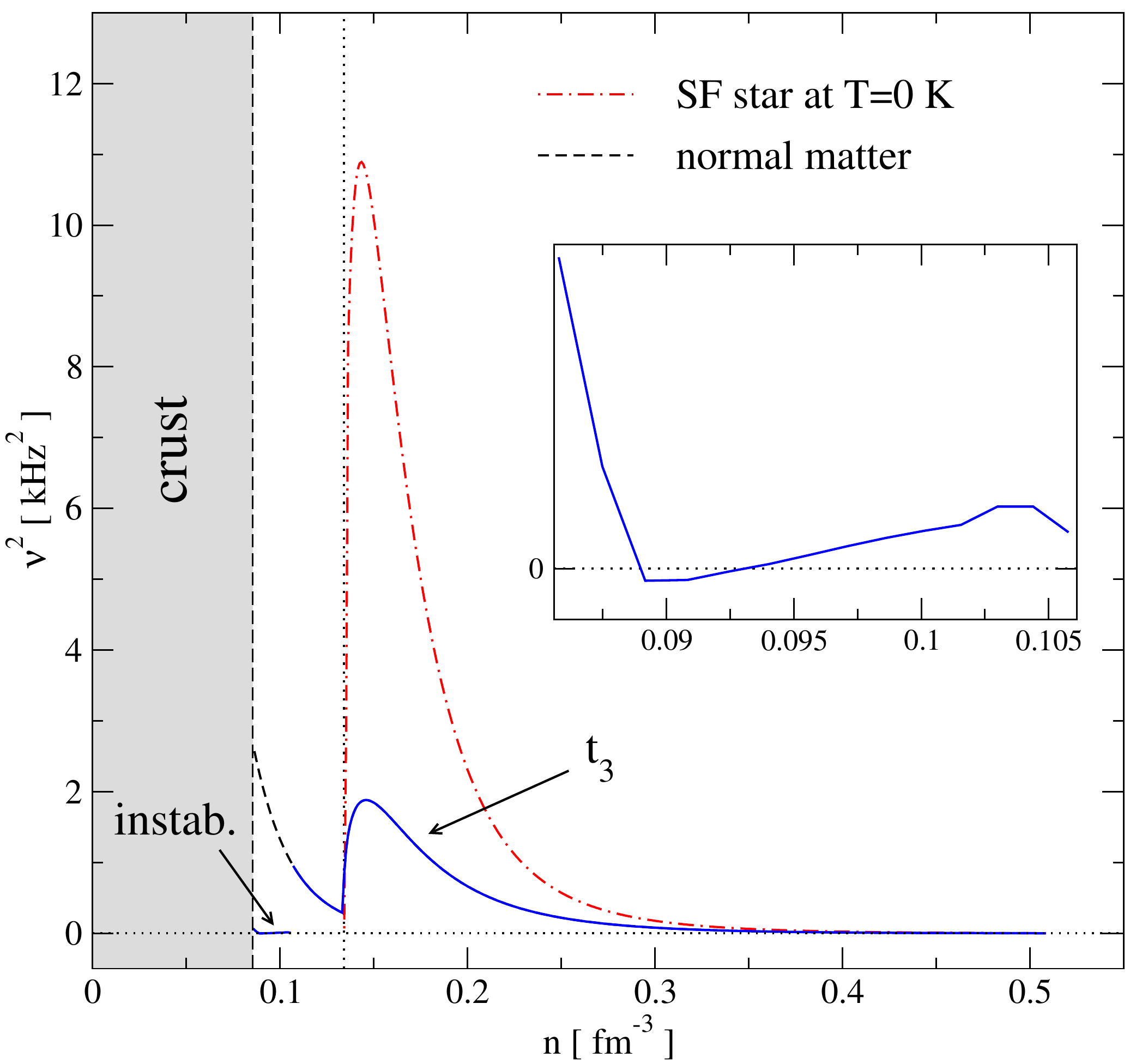}  
\caption{An illustration of the presence of a region of unstable g-modes. This example corresponds to the single one of our extracted temperature distributions for which an instability is present, at $t_3=10^{-2}$~yr (cf. Figure~\ref{fig:cooling}). We  see that the instability is weak  and strongly localised, meaning that only very short wavelength motion would actually be unstable. 
\label{instab}}
\end{center}
\end{figure*}

Finally, let us consider the suggestion from \citet{gusk1} that the superfluid g-modes may become unstable and trigger a convective phase in a young neutron star. From the results we have discussed so far it is clear that, if such an instability is to occur it has to be located in the outer part of the crust where the muons are absent. In essence, we need the overall sign of the terms in the bracket of \eqref{entro} to be negative for this instability to be present. Considering this issue for our chosen equation of state and the single stellar model we are focussing on, we have carefully checked the various temperature distributions extracted from our cooling simulation. From this data we have found one single case where an instability is present. This case is illustrated in Figure~\ref{instab}. We see that the instability is weak ($\sigma^2_S$ only becomes marginally negative) and strongly localised, meaning that only very short wavelength motion would actually be unstable. The instability is present after about a day of thermal evolution (which is long enough that the artificial isothermal initial temperature distribution has filtered through the system, as this happens on a timescale of seconds), but it is gone well before the system is one year old. This seems to suggest that this convective instability is unlikely to play much of a role in the evolution of a neutron star. Having said that, it is conceptually interesting and one should keep in mind that we have only considered one particular model.

\subsection{Entrainment}

If we want to account for the relevant physics of a neutron star core, it is imperative that we include the entrainment effect in our model. 
Entrainment is known to play an important role in superfluid dynamics \citep{mendell,prix,crust}. This is natural as it encodes how easy (or hard) it is for a superfluid component to move relative to the other parts of the system. It is a non-dissipative effect, which essentially represents the effective mass of the superfluid component. There are two, formally different but de facto equivalent, ways of incorporating the entrainment. The first approach encodes the effect in a mass-density matrix $\rho_{\x\y}$ which designates how much one component tends to flow along with the other(s) \citep{mendell}. An alternative, which is more natural from the relativistic point-of-view \citep{lrr-2007}, is to account for the effect through the momentum of each component. In this approach, the entrainment leads to the momentum of a given component $\mu^\x_a$ no longer being proportional to the corresponding flow $n_\x^a$, but rather a linear combination of all the flows that are entrained.

%-----------------------------FIG. 7------------------------------------------%
%------------------------------------------------------------------------------%

Under the conditions that prevail in a cold neutron star core, the main cause of entrainment is the strong interaction. As a superfluid neutron moves, a virtual cloud of entrained protons tries to move with it (and vice versa). The effect does not depend (at least not much) on the temperature. In the following we quantify this strong interaction entrainment by a coefficient $\mathcal A^{\n\p}$. In a finite-temperature superfluid, one also has to account for the presence of thermal excitations. As argued by \citet{and13}, the dynamical role of the excitations can be accounted for in terms of entrainment between the superfluid and the heat \citep{2011RSPSA.467..738L,2011CQGra..28s5023A}. In the following, we represent this effect by a coefficient $\mathcal A^{\n\s}$. The upshot is that the momentum of the superfluid neutrons in our mixture is  given by 
\citetext{see \citealp{and13} for  a detailed discussion}
%(see \citet{and13} for  a detailed discussion)
\begin{equation}
\mu^\n_a = \mathcal B^\n n^\n_a +  \mathcal{A}^{\n \p}  n^\p_a+  \mathcal{A}^{\n \s} s_a \ ,
\end{equation}
where the three coefficients $\mathcal B^\n$, $\mathcal{A}^{\n \p}$ and $\mathcal{A}^{\n \s}$ follow from the equation of state.
It is worth noting that we have assumed that the protons, electrons and muons all flow together. Projecting the neutron momentum along the neutron flow, we see that the corresponding chemical potential is given by
\begin{equation}
\mu_\n = n_\n \mathcal{B}^{\n} + n_\p \mathcal{A}^{\n \p} + s \mathcal{A}^{\n \s} \ .
\end{equation}
This expression is the key to adding the entrainment to our previous analysis of the g-mode problem. This extension turns out to be surprisingly straightforward.
After consulting the discussion from the previous Section, we simply introduce a new parameter
\begin{equation}
\beta = \frac{ n_\n \mathcal{B}^{\n}}{ \mu_\n} = 1- {1\over \mu_\n} \left( n_\p \mathcal{A}^{\n \p} + s \mathcal{A}^{\n \s}\right) \ .
\end{equation}
As discussed in Appendix~B, there are two key limits to consider. First of all, in the limit of zero entrainment we have $\beta = 1$. Meanwhile,  $\beta$ will diverge as $T\to T_{\ch \n}$, as we approach the critical density for onset of neutron superfluidity (see Appendix~B, especially Eq.~\eqref{B13}).

The analysis leading to the dispersion relation now proceeds along the same steps as in the case without entrainment. We only have to account for the fact that the momenta are linear combinations of the two fluxes and (carefully) redefine some of the quantities involved. We will also need slightly different combinations of the various thermodynamical derivatives. The logic of the analysis is, however, the same as before. Hence, we  relegate the details to Appendix~B and  focus on the results here. The main question is how the entrainment impacts on the local wave propagation.

Let us first consider the sound waves. In the high-frequency limit, we find that the dispersion relation \eqref{entdisp} reduces to
\begin{equation}
   \left(  \mathcal L_b^2 - \sigma^2 \right) \left({\mathcal{L}_{\n}^2 - \sigma^2 }\right) -  {\bar\Gamma_1^b  \over \bar \Gamma_\mu^b}  {G_\mu \over G_p} \sigma^4 \approx 0 \ ,
\end{equation}
where
\begin{equation}
\frac{1  }{G_p} =  -   \frac{\beta - x_\n}{x_\p }   \left(   \frac{\beta - 1 }{\beta - x_\n}    \frac{1}{\Gamma_1^b} - \frac{1}{\Gamma_1^\n} \right)   \, , 
\end{equation}
and
\begin{equation}
\frac{1  }{G_\mu} =  -   \frac{\beta - x_\n}{x_\p }   \left(   \frac{\beta - 1 }{\beta - x_\n}    \frac{1}{\Gamma_\mu^b} - \frac{1}{\Gamma_\mu^\n} \right)   \, . \\
\end{equation}
In these expressions, $\mathcal L_b^2$ takes the same form as in \eqref{Lb2}, but we now have 
\begin{equation}
\alpha_{\ch}  =   \frac{ \beta w - n_{\n} \mu_{\n} }{ \beta - x_{\n}} \ .
\end{equation}
Meanwhile, 
\begin{equation}
\mathcal L_\n ^2 =    \frac{l(l+1) }{ r^2 } e^{2\Phi}   G_\mu   \ .
\end{equation}
%
%------------------------------FIG. 8------------------------------------------%
%------------------------------------------------------------------------------%
In general, we have the two anticipated sets of sound waves. However, the situation changes as we approach the critical temperature for the onset of superfluidity. 
In this limit, we have 
\begin{equation} \lim _{T \to T_{\ch \n} } \alpha_{\ch}  = w  \,  ,
\end{equation}
and
\begin{equation}
\lim _{T \to T_{\ch \n} } \frac{1}{G_p } \sim \lim _{T \to T_{\ch \n} } \frac{1}{G_\mu }    = \infty \,  , \\
\end{equation}
This means that 
\begin{equation}
 \mathcal L_b^2   \to     \frac{l(l+1) }{ r^2 } e^{2\Phi} \,  \frac{p   \bar \Gamma_1^b }{w }   \, , 
\label{entsound} \end{equation}
 and
 \begin{equation}
\mathcal L_\n ^2 =    \frac{l(l+1) }{ r^2 } e^{2\Phi}   G_\mu   \to 0  \, .
\end{equation}
We learn that one set of sound waves disappears as we cross the superfluid transition. This is, of course, as expected. The ``second sound'' is not supported in normal matter. As far as we are aware, this is the first time that this limit has been demonstrated explicitly. In particular, our analysis highlights the key role that the thermal entrainment plays in  the problem. A closer analysis of the remaining sound wave, represented by \eqref{entsound}, shows that it is close to, but not identical to, the normal sound ($\bar \Gamma_1^b$ is not identical to $\Gamma_1^b$). The typical difference is at the few percent level, depending on where the superfluid transition takes place in the core. In other words, a small discontinuity remains in the local sound-wave frequency as we cross the critical density between superfluid and normal regions.
%In the limit, we are left with a single set of sound waves given by
%\begin{equation}
%\sigma^2 \approx \left( 1 - {1-\bar \Gamma_1^b/\bar \Gamma_1^\n\over 1- \bar \Gamma_\mu^b/\bar \Gamma_\mu^\n} \right)^{-1} \mathcal L_b^2
%\end{equation}
%\comment{How close to normal fluid sound? Can we prove something?}

%------------------------------FIG. 9------------------------------------------%
\begin{figure*}
\begin{center}
\includegraphics[height=75mm]{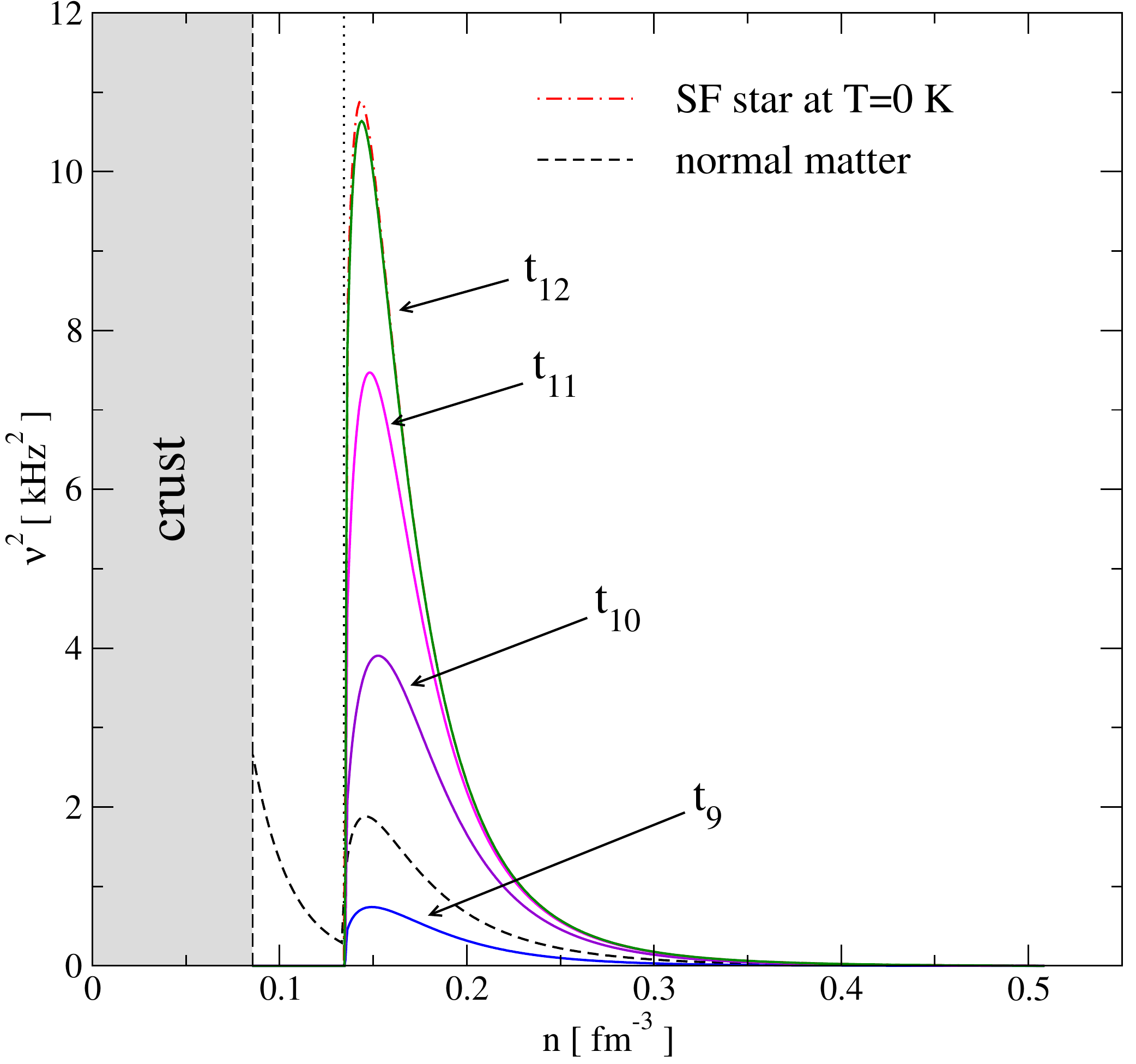}  
\caption{An illustration of the effect that  thermal excitations (accounted for through the entrainment between superfluid and entropy) has on the transition between different layers in the star. The superfluid result starts out significantly below the normal-fluid g-mode frequency at the superfluid transition temperature 
\citetext{which accords with the results of \citealp{gusk2}} 
%(which accords with the results of \citet{gusk2}) 
and then rises towards the cold superfluid result as the star ages. The results relate to stars aged $t_9=10^2$~yr, $t_{10}=10^3$~yr, $t_{11}=10^4$~yr and $t_{12}=10^5$~yr, respectively, cf. the cooling curves  in Figure~\ref{fig:cooling}. A key take-home message is that the oscillation spectrum may still be evolving even for rather mature stars.
\label{entrain}}
\end{center}
\end{figure*}
%------------------------------------------------------------------------------%

Turning to the low-frequency regime, we find that \eqref{entdisp} leads to the presence of g-modes with frequency
\begin{equation}
\sigma^2 \approx \mathcal N_c^2 \ ,
 \end{equation}
where
\begin{equation}
\mathcal N_c ^2 =   \frac{ \beta \veps }{\alpha_{\ch}} N_c^2 e^{2(\Phi-\Lambda)}  =   - \frac{ \beta \veps }{\alpha_{\ch}}  \frac{g}{\beta - x_{\n} } \sum_X  \frac{X^{\prime} }{\bar \Gamma_{X}}  e^{2(\Phi-\Lambda)} \ ,
\end{equation}
which leads to, as we approach the superfluid transition;
\begin{equation}  
\mathcal N_c ^2 \to - \frac{ g  \veps }{w }   \sum_X  \frac{X^{\prime} }{\bar \Gamma_{X}}  e^{2(\Phi-\Lambda)} \quad \mbox{as } T\to T_{\ch \n} \, .
\end{equation}
As in the case of the sound waves, this result does not limit to the normal-fluid g-mode frequency at the superfluid transition. In fact, the discontinuity in this case is significant. The superfluid results starts out significantly below the normal-fluid g-mode frequency at the superfluid transition temperature 
\citetext{which accords with the results of \citealp{gusk2}} 
%(which accords with the results of \citet{gusk2}) 
and then rises towards the cold superfluid result as the star ages. 
Making use of our cooling data, we obtain the results shown in Figure~\ref{entrain}. This figure illustrates the effect that  thermal excitations (accounted for through the entrainment between superfluid and entropy) has on the transition between different layers in the star. Note that we have not included the entrainment due to the strong interaction as is it essentially temperature independent (and it is a small effect). This demonstration shows that it is important to account for the thermal effects when considering the dynamics of superfluid neutron stars, and emphasises the need for more detailed modelling building on, for example, \citet{gusa,2011PhRvD..83j3008K,2011MNRAS.418L..54C,and13,2014PhRvD..90b4010G}.

\section{Concluding remarks}

In this paper we have investigated the local dynamics of a realistic neutron star core, accounting for composition gradients, superfluidity and thermal effects. Our main focus was on the gravity g-modes, which are supported by composition stratification and thermal gradients, but we also provided results relevant for the sound waves of the system. We have derived the detailed equations that govern the problem, paying particular attention to how the physics that is encoded in the equation of state enters and distinguishing between normal and superfluid regions. The analysis highlighted a number of  issues that need to be kept in mind whenever equation of state data is compiled from nuclear physics calculations for use in this kind of analysis. We provided explicit results for a particular stellar model and a specific equation of state 
\citetext{BSk20 from \citealp{bsk2}}, 
%(BSk20 from \citet{bsk2}), 
making use of accurate cooling simulations to show how the local wave spectrum evolves as the star ages from adolescence to maturity. Our results confirm the expectation that the composition gradient is dominated by the muons beyond the density at which they first appear \citep{gusk2}. At lower densities, where there are no muons, the support for the g-modes is entirely thermal once the star cools below the superfluid transition. We confirm the recent suggestion that the g-modes in this region may be unstable \citep{gusk1}, but our results indicate that this instability will be weak and would only be present for a short period of the star's life (less than a year). A novel technical aspect of our analysis is associated with the thermal excitations at finite temperatures which are accounted for in terms of entrainment between the entropy and the superfluid component. Once the thermal entrainment is accounted for we see that the oscillation spectrum may still be evolving even for rather mature stars. Finally, we discussed the difference between the normal sound waves and the second sound that is present only in the superfluid regions. 

This work extends the state-of-the-art in several ways. Most importantly, we have shown how detailed equations of state data can be implemented to study neutron star dynamics at a serious level of realism. This poses two main challenges for the future. First of all, we need to build on this work to construct global models for dynamical neutron stars, as required for detailed asteroseismology studies. Secondly, we need a continued dialogue with nuclear physics experts both to expand the set of equation of state models that provide all the information required for this kind of study, and also to complete the models we have used in this work (e.g. concerning consistent superfluid pairing gaps).

\section*{Acknowledgements}

A.P. acknowledges support from the European Union Seventh
Framework Programme (FP7/2007-2013) under grant
agreement n$^{\rm o}$ 267251 ``Astronomy Fellowships in Italy (AstroFIt)''. The work of N.A. and W.C.G.H. is supported by the STFC in the UK. 

\appendix
%%%%%%%%%%%%%%%
\section{Required thermodynamical relations}
%%%%%%%%%%%%%%%

Let us now take a closer look at how we  evaluate the different thermodynamical partial derivatives we need. This issue is not trivial as our perturbative analysis made use of variables that require a somewhat unusual description of the equation of state. Basically, it is important to keep track of what is being held fixed when the various adiabatic indices are evaluated. As a first step, which takes us a bit closer to where we need to be, let us assume that the equation of state is given in terms of the variables
$n=n_\n+n_\p$, $x_\e = n_\e/n$, $M=n_\mu/n_\e$ and $S=s/n_\e$. 
As our main interest is in the g-modes, let us  focus on the quantities we need to determine them. The other thermodynamical quantities can be obtained in an analogous fashion. 

In terms of the chosen variables, the energy variation can be written
\begin{equation}
d\veps = {p+\veps\over n} dn + nST dx_\e + n x_\e T dS \ ,
\end{equation}
(where we have imposed the relevant equilibrium conditions to simplify the prefactors).

\subsection{The muon contribution}

Let us, first of all, work out the contribution due to the muon gradient. This means assuming that $S$ is held fixed, so we take $dS=0$ in the following. 
We want to work out
\begin{equation}
{\veps \over \bar \Gamma_M} = \left( {\partial \veps \over \partial M}\right)_{p,\mu_\n,S}  \ .
\end{equation}
As this involves holding the pressure constant, we  use  the definition of the pressure to get
\begin{equation}
dn = - a_1 dx_\e - a_2 dM \ ,
\end{equation}
where
\begin{equation}
a_1 = \left({\partial p \over \partial n}\right)^{-1} \left({\partial p \over \partial x_\e}\right)\ ,
\end{equation}
(here and in the following we need to keep in mind that the other variables are to be held fixed when the partial derivatives are evaluated) and
\begin{equation}
a_2 = \left({\partial p \over \partial n}\right)^{-1} \left({\partial p \over \partial M}\right) = x_\e \left({\partial p \over \partial n}\right)^{-1} \left({\partial p \over \partial x_\mu}\right)  \ .
\end{equation}
The last equality follows since $n$ and $x_\e$ are held fixed, which means that $dM=(1/x_\e)dx_\mu$.
Next, the requirement that $\mu_\n$ is constant leads to 
the relation
\begin{equation}
dx_\e = - a_3 dM\ ,
\end{equation}
where
\begin{equation}
a_3 =  \left[ a_1 - \left({\partial \mu_\n \over \partial n}\right)^{-1} \left({\partial \mu_\n \over \partial x_\e}\right) \right]^{-1} \left[ a_2 - \left({\partial \mu_\n \over \partial n}\right)^{-1} \left({\partial \mu_\n \over \partial M}\right) \right] \ .
\end{equation}
By combining these results, we arrive at the expression we need
\begin{equation}
\left( {\partial \veps \over \partial M}\right) = {p+\veps \over n} (a_1 a_3 - a_2) - nST a_3 \ .
\end{equation}

\subsection{The entropy contribution}

Turning to the thermal contribution, we instead hold $M$ fixed and
work out
\begin{equation}
{\veps \over \bar \Gamma_S} = \left( {\partial \veps \over \partial S}\right)_{p,\mu_\n,M}  \ .
\end{equation}
For obvious reasons, the results are similar to those for the muons. Holding the pressure fixed leads to 
\begin{equation}
dn = - b_1 dx_\e - b_2 dS \ ,
\end{equation}
where
\begin{equation}
b_1 = \left({\partial p \over \partial n}\right)^{-1} \left({\partial p \over \partial x_\e}\right)\ ,
\end{equation}
(the only difference from the previous case is that $M$ is held fixed rather than $S$) and
\begin{equation}
b_2 = \left({\partial p \over \partial n}\right)^{-1} \left({\partial p \over \partial S}\right)\ .
\end{equation}
Next, keeping  $\mu_\n$ fixed leads to 
the relation
\begin{equation}
dx_\e = - b_3 dS\ ,
\end{equation}
where
\begin{equation}
b_3 =  \left[ b_1 - \left({\partial \mu_\n \over \partial n}\right)^{-1} \left({\partial \mu_\n \over \partial x_\e}\right) \right]^{-1} \left[ b_2 - \left({\partial \mu_\n \over \partial n}\right)^{-1} \left({\partial \mu_\n \over \partial S}\right) \right] \ .
\end{equation}
By combining these results, we arrive at 
\begin{equation}
\left( {\partial \veps \over \partial S}\right) = {p+\veps \over n} (b_1 b_3 - b_2) - nST b_3 + n x_\e T \ .
\end{equation}

In the model we consider in this paper, the thermal contributions are added perturbatively to a cold background star. Moreover, it tends to be the case that the thermal effects can be neglected compared to the cold parameters. For example,  for the range of temperatures we consider, the thermal pressure can  always be neglected compared to the cold degeneracy pressure throughout the star's core. For our model, we find that the thermal pressure leads to
\begin{equation}
\left( {\partial p_{th} \over \partial S} \right)_{n,x_\e,M} = {2 p_{th} \over S} = {n_\e T \over 3} \ .
\end{equation}
This provides us with the information required to evaluate the $b_2$ coefficient above. Meanwhile, for the other partial derivatives, we can neglect the thermal pressure. Writing the total pressure as $p=p_0+p_{th}$ we then have
\begin{equation}
\left({\partial p \over \partial n}\right) = \left({\partial p_0 \over \partial n}\right) \ , \quad \mbox{ and } \quad \left({\partial p \over \partial x_\e}\right) = \left({\partial p_0 \over \partial x_\e}\right) \ .
\end{equation}

Turning to the neutron chemical potential, we first of all have
\begin{equation}
\left( {\partial \mu_\n \over \partial S} \right)_{n,x_\e,M} = 0 \ .
\end{equation}
The other partial derivatives we need can be accurately obtained from the cold equation of state. 

Combining the results, we have the final relation
\begin{multline}
\left( {\partial \veps \over \partial S}\right)_{p,\mu_\n,M}  = {n_\e T \over 3} \left({\partial p_0 \over \partial n}\right)^{-1} \left[ {p+\veps \over n} \left({\partial p_0 \over \partial n}\right)^{-1}\left({\partial p_0 \over \partial x_\e}\right) - nST \right] \left[  \left({\partial p_0 \over \partial n}\right)^{-1}\left({\partial p_0 \over \partial x_\e}\right) -  \left({\partial \mu_\n \over \partial n}\right)^{-1}\left({\partial \mu_\n\over \partial x_\e}\right)  \right]^{-1} \\
+ n_\e T \left[ 1 - {p+\veps \over n}  \left({\partial p_0 \over \partial n}\right)^{-1} \right] \ ,
\end{multline}
where we note that the overall factor of $T$ implies that the thermal support for the g-modes will decrease as the star cools. This is, obviously, as expected.

In practice, we can use
\begin{multline}
\left( {\partial \veps \over \partial S}\right)_{p,\mu_\n,M}  = {n_\e T \over 3} \left({\partial p_0 \over \partial n}\right)^{-1} \left[ {p+\veps \over n} \left({\partial p_0 \over \partial n}\right)^{-1}\left({\partial p_0 \over \partial x_\e}\right) \right] \left[  \left({\partial p_0 \over \partial n}\right)^{-1}\left({\partial p_0 \over \partial x_\e}\right) -  \left({\partial \mu_\n \over \partial n}\right)^{-1}\left({\partial \mu_\n\over \partial x_\e}\right)  \right]^{-1} \\
+ n_\e T \left[ 1 - {p+\veps \over n}  \left({\partial p_0 \over \partial n}\right)^{-1} \right] \ ,
\end{multline}
where the terms in the brackets can be obtained from the cold equation of state.

\subsection{Other thermodynamical relations}

In addition to the thermodynamical derivatives required to determine the g-modes, we need a more complete set if we are to solve the full dispersion relation. These involve;
\begin{equation}
c_b^2 = {p\bar\Gamma_1^b \over n} =  \left( {\partial p \over \partial n }\right)_{\mu_\n,X} = {\partial p \over\partial x_\e} \left[\left({\partial p \over \partial x_\e}\right)^{-1} {\partial p \over \partial n}- \left({\partial \mu_\n \over \partial x_\e}\right)^{-1} {\partial \mu_\n \over \partial n} \right] \ ,
\end{equation}

\begin{equation}
\bar c_s^2 =  \left( {\partial p \over \partial \veps }\right)_{\mu_\n,X} =  \left[ {p+\veps \over n} - nST  \left( {\partial \mu_\n \over \partial x_\e}\right)^{-1}  \left( {\partial \mu_\n \over \partial n}\right) \right]^{-1} c_b^2 \approx {n \over p+\veps } c_b^2\ ,
\end{equation}

\begin{multline}
\bar \Gamma^\n_\mu  = {n_\n \over \mu_\n} \left( {\partial \mu_\n \over \partial n_\n}\right)_{p,X} =  {n_\n \over \mu_\n} {\partial \mu_\n \over\partial x_\e}  \left[ x_\n + n(1+M) \left({\partial p \over \partial x_\e}\right)^{-1} {\partial p \over \partial n} \right]^{-1} \left[\left({\partial \mu_\n \over \partial x_\e}\right)^{-1} {\partial \mu_\n \over \partial n} - \left({\partial p \over \partial x_\e}\right)^{-1} {\partial p \over \partial n}\right] \\
=  - {n_\n \over \mu_\n} {\partial \mu_\n \over\partial x_\e} \left( {\partial p \over \partial x_\e} \right)^{-1} c_b^2 \left[ x_\n + n(1+M) \left({\partial p \over \partial x_\e}\right)^{-1} {\partial p \over \partial n} \right]^{-1} \ ,
\end{multline}

\begin{equation}
\bar \Gamma^b_\mu = - {n \over \mu_\n} {\partial \mu_n \over \partial x_\e} \left[\left({\partial p \over \partial x_\e}\right)^{-1} {\partial p \over \partial n}- \left({\partial \mu_\n \over \partial x_\e}\right)^{-1} {\partial \mu_\n \over \partial n} \right] = - {n\over \mu_\n } {\partial \mu_n \over \partial x_\e} \left({\partial p \over \partial x_\e}\right)^{-1} c_b^2\ ,
\end{equation}

\begin{multline}
{1 \over \bar \Gamma_\mu} = {\mu_\n \over \veps}  \left( {\partial \veps \over \partial \mu_\n}\right)_{p,X} \\
= {\mu_\n \over \veps} \left( {\partial \mu_\n \over \partial x_\e}\right)^{-1} 
 \left[\left({\partial \mu_\n \over \partial x_\e}\right)^{-1} {\partial \mu_\n \over \partial n} - \left({\partial p \over \partial x_\e}\right)^{-1} {\partial p \over \partial n}\right]^{-1}
\left[ {p+\veps \over n} - nST  \left( {\partial \mu_\n \over \partial x_\e}\right)^{-1}  \left( {\partial \mu_\n \over \partial n}\right) \right] \\
= - {1\over \bar c_s^2}  {\mu_\n \over \veps} \left( {\partial \mu_\n \over \partial x_\e}\right)^{-1} \left( {\partial p \over \partial x_\e}\right)\ ,
\end{multline}

\begin{equation}
{1 \over \bar \Gamma_1^\n} = {p\over n} \left( {\partial n_\n \over \partial p} \right)_{\mu_\n, X} =  {p\over n} \left[ x_\n + n(1+M) \left({\partial \mu_\n \over \partial x_\e}\right)^{-1} {\partial \mu_\n \over \partial n} \right]  {1 \over c_b^2}\ ,
\end{equation}

\begin{multline}
{1 \over \bar \Gamma^\n_M} = {1\over n_\n} \left( {\partial n_\n \over \partial M} \right)_{p,\mu_\n,S} =  {1\over n_\n (1+M)} \left[ \left({\partial \mu_\n \over \partial x_\e}\right)^{-1} {\partial \mu_\n \over \partial n} -  \left({\partial p \over \partial x_\e}\right)^{-1} {\partial p \over \partial n}  \right]^{-1} \\
\times \Bigg\{ \left[ x_\e - (1+M)  \left({\partial p \over \partial x_\e}\right)^{-1} {\partial p \over \partial M}\right] \left[ x_\n + n(1+M)  \left({\partial \mu_\n \over \partial x_\e}\right)^{-1} {\partial \mu_\n \over \partial n} \right]\\
- \left[ x_\e - (1+M)  \left({\partial \mu_\n \over \partial x_\e}\right)^{-1} {\partial \mu_\n \over \partial M}\right] \left[ x_\n + n(1+M)  \left({\partial p \over \partial x_\e}\right)^{-1} {\partial p \over \partial n} \right] \Bigg\}\ ,
\end{multline}

\begin{multline}
{1 \over \bar \Gamma^\n_S} =  {1\over n_\n} \left( {\partial n_\n \over \partial S} \right)_{p,\mu_\n,M} =  {1\over n_\n } \left[ \left({\partial \mu_\n \over \partial x_\e}\right)^{-1} {\partial \mu_\n \over \partial n} -  \left({\partial p \over \partial x_\e}\right)^{-1} {\partial p \over \partial n}  \right]^{-1} \\
\times \Bigg\{  \left({\partial p \over \partial x_\e}\right)^{-1} {\partial p \over \partial S} \left[ x_\n + n(1+M)  \left({\partial \mu_\n \over \partial x_\e}\right)^{-1} {\partial \mu_\n \over \partial n} \right] 
- \left({\partial \mu_\n \over \partial x_\e}\right)^{-1} {\partial \mu_\n \over \partial S} \left[ x_\n + n(1+M)  \left({\partial p \over \partial x_\e}\right)^{-1} {\partial p \over \partial n} \right] \Bigg\}
\\
= - {1\over n_\n c_b^2 }  {\partial p \over \partial S} \left[ x_\n + n(1+M)  \left({\partial \mu_\n \over \partial x_\e}\right)^{-1} {\partial \mu_\n \over \partial n} \right] \ ,
\end{multline}

\begin{multline}
{1 \over \bar \Gamma^b_M } = {1\over n} \left( {\partial n\over \partial M} \right)_{p,\mu_\n,S} =  - {1\over n} \left[\left({\partial \mu_\n \over \partial x_\e}\right)^{-1} {\partial \mu_\n \over \partial n} - \left({\partial p \over \partial x_\e}\right)^{-1} {\partial p \over \partial n} \right]^{-1} \left[ \left({\partial \mu_\n \over \partial x_\e}\right)^{-1} {\partial \mu_\n \over \partial M} - \left({\partial p \over \partial x_\e}\right)^{-1} {\partial p \over \partial M}\right] \\
=  { 1 \over c_b^2} {\partial p \over \partial x_\e} \left[ \left({\partial \mu_\n \over \partial x_\e}\right)^{-1} {\partial \mu_\n \over \partial M} - \left({\partial p \over \partial x_\e}\right)^{-1} {\partial p \over \partial M}\right] \ ,
\end{multline}

\begin{multline}
{1 \over \bar \Gamma^b_S } = {1\over n} \left( {\partial n\over \partial S} \right)_{p,\mu_\n,M} =  - {1\over n} \left[\left({\partial \mu_\n \over \partial x_\e}\right)^{-1} {\partial \mu_\n \over \partial n} - \left({\partial p \over \partial x_\e}\right)^{-1} {\partial p \over \partial n} \right]^{-1} \left[ \left({\partial \mu_\n \over \partial x_\e}\right)^{-1} {\partial \mu_\n \over \partial S} - \left({\partial p \over \partial x_\e}\right)^{-1} {\partial p \over \partial S}\right] \\
=  { 1 \over c_b^2} {\partial p \over \partial x_\e} \left[ \left({\partial \mu_\n \over \partial x_\e}\right)^{-1} {\partial \mu_\n \over \partial S} - \left({\partial p \over \partial x_\e}\right)^{-1} {\partial p \over \partial S}\right] 
=  - { 1 \over c_b^2}  {\partial p \over \partial S}\ .
\end{multline}

It is worth keeping in mind that, for the model we are using, we have
\begin{equation}
{\partial p \over\partial S} = {n_\e T \over 3} \ .
\end{equation}

\section{Accounting for entrainment}

\subsection{Entrainment coefficients}

Following \citet{and13}, we encode  the entrainment between the superfluid neutrons and  thermal excitations in terms of a coefficient $\mathcal A^{\n\s}$ which can be obtained from the results of  \citet{gushae}. They  provide a (non-relativistic) mass density matrix, given by;
\begin{align}
& \rho_{\n \n } =  \left( 1 - f_\n \right)  \bar \rho_{\n \n} \, , \\
& \rho_{\p \p } =  \left( 1 - f_\p \right)  \bar \rho_{\p \p} \, , \\
& \rho_{\n \p } =  \left( 1 - f_\n \right) \left( 1 - f_\p \right) \bar \rho_{\n \p} \, , 
\end{align}
where $f_\x$ are  temperature-dependent functions that tend to zero in the $T\to 0$ limit and approach unity when  $T\to T_{\ch \x}$. The particular form for these functions is given by \citet{gnedin}. In the above expression, the $\bar \rho_{\x\y}$ quantities  depend only weakly on the temperature. Hence, we take them to be given by the 
zero temperature expressions of the mass density matrix in the following. Expressing the results in terms of the effective proton mass $m_\p^\ast$ we then have \citep{and13}
\begin{align}
& \bar \rho_{\n \n } =  \rho_\n  m_\p ^\ast   \left[  m_\p ^\ast - \frac{n_\p }{n_{\n}} \left(  m - m_\p ^\ast \right) \right]  ^{-1}  \approx \rho_\n  \, , \label{rhox1}\\
& \bar \rho_{\p \p } =  \rho_\p  \left[  m - \frac{n_\p }{n_{\n}} \left(  m - m_\p ^\ast \right) \right]   \left[  m_\p ^\ast - \frac{n_\p }{n_{\n}} \left(  m - m_\p ^\ast \right) \right]  ^{-1}  \approx \rho_\p  \frac{m}{m_\p^\ast}\, , \\
& \bar  \rho_{\n \p } =  - \rho_\p \left( m -  m_\p ^\ast  \right)  \left[  m_\p ^\ast - \frac{n_\p }{n_{\n}} \left(  m - m_\p ^\ast \right) \right]  ^{-1}  \approx   - \rho_\p \left( \frac{m - m_\p ^\ast } { m_\p ^\ast  } \right) \label{rhox2} \, ,
\end{align}
 where $\rho_\x = m n_\x$ is the non-relativistic mass density and the approximations are valid when $n_\p / n_\n \ll 1 $.

The translation to our relativistic model has already been discussed by \citet{and13}. It proceeds in two steps. First we identify the correspondence with the leading order 
Newtonian terms;
\begin{align}
&  \mathcal{B} ^{\n } \approx \tilde{ \mathcal{K} }^{\n \n } =  \frac{m^2 \bar \rho_{\p \p } }{\left( 1 - f_\n \right) \mathcal{R} } \, , \\
&  \mathcal{A} ^{\n \p } \approx   \tilde{ \mathcal{K} }^{\n \p } =  -  \frac{m^2 \bar \rho_{\n \p } }{ \mathcal{R} } \, , \\
& \mathcal{A}^{\n \s} =  \frac{m}{s} \left[  1 - \frac{\bar \rho_{\p\p}  \rho_\n - \left( 1 - f_\n \right) \bar \rho_{\n\p} \rho_\p }{ \left( 1 - f_\n \right) \mathcal{R}} \right]\ ,
\end{align}
 where we have defined
\begin{equation}
 \mathcal{R} = \bar \rho_{\n \n } \bar \rho_{\p \p } - \left( 1 - f_\n \right) \left( 1 - f_\p \right)  \bar \rho_{\n\p} ^ 2\ .
\end{equation}
In the relativistic case, we also need to add the Newtonian chemical potential $\mu_\n^N$ to $n_\n\mathcal B^\n$. This term is (within our approximations) temperature independent. 
We then have
\begin{equation}
\mu_\n = n_n \tilde K^{\n\n} + n_\p \tilde K^{\n\p} + s \mathcal A^{\s\n} +\mu_\n^N = m+\mu_\n^N\ .
\end{equation}
The last equality shows that the neutron chemical potential is not affected by the presence of thermal excitations. 

In the cold limit, we can use \eqref{rhox1}--\eqref{rhox2} (not the approximations, though!) to show that 
\begin{equation}
\beta \to 1 + {n_\p \over \mu_\n} {m^2 \rho_{\n\p} \over \mathrm{det} \rho}\ ,
\end{equation}
as expected.
Meanwhile, as the critical temperature is approached we have the leading order behaviour
\begin{equation}
\beta \approx {m\over \mu_\n} {\rho_\n \rho_{\p\p} \over (1 - f_\n) \mathcal R}\ .
\label{B13}
\end{equation}
As we will see, 
the dynamics of the system is controlled by  the divergence of this quantity as $T \to T_{\ch \n}$. 

Finally, it is worth pointing out that, if we ignore the thermal effects then the strong interaction contributes to the entrainment exactly as in a cold superfluid neutron star. We would have
\begin{equation}
\tilde K^{\n\p} = {1\over n_\n} \left( m-m_\p^\ast \right) \ .
\end{equation}
Although important, this effect is relatively small. As our main focus is on the thermal contributions we have not accounted for it in the results shown in Figure~\ref{entrain}.

\subsection{Dispersion relation (with entrainment)}

When we turn to the perturbation problem, we see that the entrainment requires a slight redefinition of the displacement vectors. Specifically, we have 
\begin{equation}
 e^{-\Phi} \partial_{t} \xi^{a }     = \delta u^{a} +  x_{\n} \delta v^{a} \, ,
 \end{equation}
 and
 \begin{equation}
e^{-\Phi}  \partial_{t} \eta^{a }  = \delta u^{a} +  \beta \delta v^{a} \ . 
\end{equation}

The radial components of the momentum equations are then given by
\begin{align}
&  \sigma^2  \left( \alpha_{\ch}   W  - \alpha_{\n}    W_{\n}  \right) e^{-2(\Phi-\Lambda)}  =  \partial_{r} \hat p   +   g  \hat w  \,  ,  \label{eq:M1b} \\
&  \sigma^2   W_{\n}   e^{-2(\Phi-\Lambda)}  =   \partial_{r} \hat \mu_{\n}^{\star} \, ,  
\end{align} 
while the angular parts are
\begin{align}
&  \sigma^2  \left( \alpha_{\ch}   V  - \alpha_{\n}    V_{\n}  \right) e^{-2 \Phi}  =  \hat p   \,  ,  \\
&  \sigma^2   V_{\n}   e^{-2 \Phi}  =  \hat \mu_{\n}^{\star} \, .
\end{align} 
In the previous expressions, we have defined the following quantities:
\begin{align} 
& \alpha_{\ch}  =   \frac{ \beta w - n_{\n} \mu_{\n} }{ \beta - x_{\n}}  \, ,  \\
& \alpha_{\n}    =   \frac{ x_{\n} w - n_{\n} \mu_{\n}}{ \beta - x_{\n}}  \ .
\end{align}

The two conservation laws lead to 
\begin{align}
& W^{\prime}+ \left[ \frac{n^{\prime}}{n}  + \frac{2}{r} + {\Lambda'} \right] W =  \frac{l(l+1) }{ r^2} V -   \frac{ \hat n }{n} \label{eq:M3} \, , \\
& \frac{x_\p }{\beta - x_\n} \left\{  W^{\prime}_{\n}+ \left[ \frac{n^{\prime}_{\n}}{n_{\n}}  + \frac{2}{r} + {\Lambda'}  + \left(  \ln  \frac{x_\p }{\beta - x_\n} \right) ^{\prime} \right] W_{\n} -  \frac{l(l+1) }{ r^2} V_{\n}  \right\} =    \frac{\beta - 1  }{\beta - x_\n} \frac{ \hat n }{n}  -  \left( \frac{\beta - 1  }{\beta - x_\n} \right) ^{\prime} W  -   \frac{ \hat n_{\n} }{n_{\n}} \label{eq:M4} \, ,
\end{align}
where we have used equation (\ref{eq:M3}) to introduce ${ \hat n }/{n}$ in equation(\ref{eq:M4}). 
The expansions of   $\delta w $,  $\delta n_\n / n_\n $ and $\delta n / n $ are similar to the zero entrainment case, the only difference is in the terms related to $x_{\x}^{\prime}$, 
as we now have:
 \begin{equation}
\delta x_{\x} =  \left(  \frac{x_\n }{\beta - x_\n}  W_\n - \frac{\beta }{\beta - x_\n} W    \right) x_{\x} ^{\prime} \ .
\end{equation}
Explicitly, we have:
\begin{equation}
\delta w = \delta \veps + \delta p = \left( 1 + {1\over \bar c_s^2} \right) \delta p + {\veps\over \bar \Gamma_\mu} \delta \mu_\n^{\star} - \frac{\veps }{ g} N_c^2 \left( x_{\n} W_{\n} - \beta W \right) \, , 
\end{equation}
\begin{equation}
 \frac{ \delta n_\n }{n_\n}  =   \frac{1}{\bar  \Gamma_1^\n } \frac{\delta p}{p} +  \frac{1}{ \bar \Gamma_\mu^\n } \delta \mu_\n^{\star} +    \left( x_{\n} W_{\n} - \beta W \right)  \mathcal A_\n  \, , 
\end{equation}
\begin{equation}
 \frac{ \delta n }{n }  
=   \frac{1}{\bar  \Gamma_1^b } \frac{\delta p}{p} +  \frac{1}{ \bar \Gamma_\mu^b } \delta \mu_\n^{\star} +    \left( x_{\n} W_{\n} - \beta W \right) \mathcal A_{b}  \, , 
 \end{equation}
where 
\begin{equation}
N_c^2  = - \frac{g}{\beta - x_{\n} } \sum_X  \frac{X^{\prime} }{\bar \Gamma_{X}}  \, ,
\end{equation}
and
 \begin{equation}
\mathcal{A} _{y} =  \frac{ 1 }{\beta - x_\n}  \frac{X ^\prime }{\Gamma_X ^\y} \ .
\end{equation}

With these definitions equation (\ref{eq:M4}) can be re-written as
\begin{align}
&  W^{\prime}_{\n}+ \left[ \frac{n^{\prime}_{\n}}{n_{\n}}  + \frac{2}{r} + {\Lambda'}  + \left(  \ln  \frac{x_\p }{\beta - x_\n} \right) ^{\prime} 
 -  x_\n   \frac{\beta - x_\n}{x_\p } \left(   \frac{\beta - 1  }{\beta - x_\n}   \mathcal{A}_{b} -  \mathcal{A}_{\n}  \right) 
\right] W_{\n} -  \frac{l(l+1) }{ r^2} V_{\n}   =    - \frac{1  }{G_p} \frac{ \hat p }{p}  -  \frac{1}{G_{\mu}} \hat \mu_\n^{\star}  +   \frac{ W}{G_{W}} \label{eq:M4a} \, ,
\end{align}
where we have defined
\begin{align}
& \frac{1  }{G_p} =  -   \frac{\beta - x_\n}{x_\p }   \left(   \frac{\beta - 1 }{\beta - x_\n}    \frac{1}{\Gamma_1^b} - \frac{1}{\Gamma_1^\n} \right)   \, , \\
& \frac{1  }{G_\mu} =  -   \frac{\beta - x_\n}{x_\p }   \left(   \frac{\beta - 1 }{\beta - x_\n}    \frac{1}{\Gamma_\mu^b} - \frac{1}{\Gamma_\mu^\n} \right)   \, , \\
& \frac{1  }{G_W} =     - \frac{\beta - x_\n}{x_\p }   \left[  \left( \frac{\beta - 1 }{\beta - x_\n} \right)^\prime + \beta \left( \frac{\beta - 1 }{\beta - x_\n}  \mathcal{A} _{b} - \mathcal{A} _{\n}     \right) \right]   \, .
\end{align}

The steps in the derivation remain exactly as in the case without entrainment. Carrying out the required algebra, we arrive at the final dispersion relation
\begin{multline}
%& 
 \left[    \tilde k^2  +    \frac{ \alpha_{\ch} }{p}  \left( \sigma^2 -  \mathcal N_c^2 \right)    \frac{  \mathcal L_b^2 - \sigma^2 }{ \sigma^2  \bar \Gamma_1^b }        \right]  \times    \left(  \tilde k ^2 +  \frac{\mathcal{L}_{\n}^2 - \sigma^2 }{G_{\mu}}   \right)  
    \\
  =  \left[   \frac{ \tilde k^2 }{G_W}  + \frac{  i k   }{G_p }   
\frac{ \alpha_{\ch} }{p}  \left( \sigma^2 -  \mathcal N_c^2 \right) \right]   \times \left[   \frac{i}{ k  \bar  \Gamma_\mu^b } \left(  \mathcal L_\star ^2  - \sigma^2  \right) 
  +   x_\n  \mathcal A_b e^{2(\Phi-\Lambda)}  \right]   \\ 
  - \frac{ i \tilde k ^2 }{k } \left(  \frac{\alpha_{\ch} x_\n}{\beta  }  \mathcal{N}_c ^2    - \alpha_\n    \sigma^2  
  + \frac{ i \sigma^2 }{ k} \frac{g \veps }{ \bar \Gamma_{\mu} }   \right) \left(  \frac{  i k   }{ p G_p }  - \frac{1}{G_W} \frac{  \mathcal L_b^2 - \sigma^2 }{ \sigma^2  p \bar \Gamma_1^b }  \right)   \ ,
% & \left( - \frac{  i k   }{G_p }  + \frac{1}{G_W} \frac{  \mathcal L_b^2 - \sigma^2 }{ \sigma^2 \bar \Gamma_1^b }  \right)  \left\{ 
%\frac{ \alpha_{\ch} }{p}  \left( \sigma^2 -  \mathcal N_c^2 \right)  \left[    \frac{  \mathcal L_\star ^2  - \sigma^2  }{  \bar  \Gamma_\mu^b }
 % - ik  x_\n  \mathcal A_b e^{2(\Phi-\Lambda)}  \right]  \right. \nn \\
%& \left.   -  \frac{ \tilde k ^2 }{p}  \left(  \frac{\alpha_{\ch} x_\n}{\beta  }  \mathcal{N}_c ^2    - \alpha_\n    \sigma^2  
%  + \frac{ i \sigma^2 }{ k} \frac{g \veps }{ \bar \Gamma_{\mu} }   \right)
%\right\}  \hat \mu_\n^{\star} 
  %   \\
  \label{entdisp}
\end{multline} 
where we have used the definitions;
\begin{equation}
\mathcal L_b^2 =      \frac{l(l+1) }{ r^2 } e^{2\Phi} \,  \frac{p  \bar  \Gamma_1^b }{\alpha_{\ch}}  \ ,
\end{equation}
\begin{equation}
\mathcal L_\n ^2 =    \frac{l(l+1) }{ r^2 } e^{2\Phi}   G_\mu   \ ,
\end{equation}
\begin{equation}
\mathcal N_c ^2 =   \frac{ \beta \veps }{\alpha_{\ch}} N_c^2 e^{2(\Phi-\Lambda)} \ ,
\end{equation}
and 
 \begin{equation}
 \mathcal{L}_{\star}^2 =   \bar  \Gamma_\mu^b \frac{l \left( l + 1 \right)}{r^2} \frac{\alpha_\n}{\alpha_\ch} e^{2 \Phi} \ .
\end{equation}
Results obtained from this dispersion relation are discussed in the main body of the paper. The key point to note from our analysis is that, even though $\beta$ diverges as we approach the critical temperature, all equations (and hence the dynamics!) remain regular.

%%%%%%%%%%%%%%%%%%%%%%%%%%%%%%%%%  BIBLIOGRAPHY  %%%%%%%%%%%%%%%%%%%%%%%%%%%%%%%
%\nocite*
%\bibliographystyle{apsrev}
\bibliographystyle{mn2e}
%\bibliography{references}

\begin{thebibliography}{}

\bibitem[\protect\citeauthoryear{{Aerts}}{{Aerts}}{2015}]{2015arXiv150306690A}
{Aerts} C.,  2015, preprint arXiv1503.06690

\bibitem[\protect\citeauthoryear{{Ainsworth}, {Wambach} \& {Pines}}{{Ainsworth}
  et~al.}{1989}]{1989PhLB..222..173A}
{Ainsworth} T.~L.,  {Wambach} J.,    {Pines} D.,  1989, Physics Letters B, 222,
  173

\bibitem[\protect\citeauthoryear{{Alford} \& {Schwenzer}}{{Alford} \&
  {Schwenzer}}{2014}]{2014PhRvL.113y1102A}
{Alford} M.~G.,  {Schwenzer} K.,  2014, Phys. Rev. Lett., 113, 251102

\bibitem[\protect\citeauthoryear{{Amundsen} \& {{\O}stgaard}}{{Amundsen} \&
  {{\O}stgaard}}{1985}]{1985NuPhA.442..163A}
{Amundsen} L.,  {{\O}stgaard} E.,  1985, Nuclear Physics A, 442, 163

\bibitem[\protect\citeauthoryear{{Andersson} \& {Comer}}{{Andersson} \&
  {Comer}}{2007}]{lrr-2007}
{Andersson} N.,  {Comer} G.,  2007, Living Reviews in Relativity, 10

\bibitem[\protect\citeauthoryear{{Andersson} \& {Comer}}{{Andersson} \&
  {Comer}}{2001}]{comer}
{Andersson} N.,  {Comer} G.~L.,  2001, MNRAS, 328, 1129

\bibitem[\protect\citeauthoryear{{Andersson}, {Ferrari}, {Jones}, {Kokkotas},
  {Krishnan}, {Read}, {Rezzolla} \& {Zink}}{{Andersson}
  et~al.}{2011}]{2011GReGr..43..409A}
{Andersson} N.,  {Ferrari} V.,  {Jones} D.~I.,  {Kokkotas} K.~D.,  {Krishnan}
  B.,  {Read} J.~S.,  {Rezzolla} L.,    {Zink} B.,  2011, General Relativity
  and Gravitation, 43, 409

\bibitem[\protect\citeauthoryear{{Andersson}, {Glampedakis}, {Ho} \&
  {Espinoza}}{{Andersson} et~al.}{2012}]{crust}
{Andersson} N.,  {Glampedakis} K.,  {Ho} W.~C.~G.,    {Espinoza} C.~M.,  2012,
  Phys. Rev. Lett., 109, 241103

\bibitem[\protect\citeauthoryear{{Andersson} \& {Kokkotas}}{{Andersson} \&
  {Kokkotas}}{2001}]{rmode}
{Andersson} N.,  {Kokkotas} K.~D.,  2001, International Journal of Modern
  Physics D, 10, 381

\bibitem[\protect\citeauthoryear{{Andersson}, {Kr{\"u}ger}, {Comer} \&
  {Samuelsson}}{{Andersson} et~al.}{2013}]{and13}
{Andersson} N.,  {Kr{\"u}ger} C.,  {Comer} G.~L.,    {Samuelsson} L.,  2013,
  Class. Quantum Grav., 30, 235025

\bibitem[\protect\citeauthoryear{{Andersson} \& {Lopez-Monsalvo}}{{Andersson}
  \& {Lopez-Monsalvo}}{2011}]{2011CQGra..28s5023A}
{Andersson} N.,  {Lopez-Monsalvo} C.~S.,  2011, Class. Quantum Grav., 28,
  195023

\bibitem[\protect\citeauthoryear{{Andersson}, {Sidery} \& {Comer}}{{Andersson}
  et~al.}{2006}]{sidery}
{Andersson} N.,  {Sidery} T.,    {Comer} G.~L.,  2006, MNRAS, 368, 162

\bibitem[\protect\citeauthoryear{{Burgio}, {Ferrari}, {Gualtieri} \&
  {Schulze}}{{Burgio} et~al.}{2011}]{2011PhRvD..84d4017B}
{Burgio} G.~F.,  {Ferrari} V.,  {Gualtieri} L.,    {Schulze} H.-J.,  2011,
  Phys. Rev. D, 84, 044017

\bibitem[\protect\citeauthoryear{{Burrows} \& {Lattimer}}{{Burrows} \&
  {Lattimer}}{1986}]{burrows}
{Burrows} A.,  {Lattimer} J.~M.,  1986, Ap. J., 307, 178

\bibitem[\protect\citeauthoryear{{Chamel}}{{Chamel}}{2008}]{chamel08}
{Chamel} N.,  2008, MNRAS, 388, 737

\bibitem[\protect\citeauthoryear{{Chen}, {Clark}, {Dav{\'e}} \&
  {Khodel}}{{Chen} et~al.}{1993}]{1993NuPhA.555...59C}
{Chen} J.~M.~C.,  {Clark} J.~W.,  {Dav{\'e}} R.~D.,    {Khodel} V.~V.,  1993,
  Nucl. Phys. A, 555, 59

\bibitem[\protect\citeauthoryear{{Chugunov} \& {Gusakov}}{{Chugunov} \&
  {Gusakov}}{2011}]{2011MNRAS.418L..54C}
{Chugunov} A.~I.,  {Gusakov} M.~E.,  2011, MNRAS, 418, L54

\bibitem[\protect\citeauthoryear{{Doneva}, {Gaertig}, {Kokkotas} \&
  {Kr{\"u}ger}}{{Doneva} et~al.}{2013}]{2013PhRvD..88d4052D}
{Doneva} D.~D.,  {Gaertig} E.,  {Kokkotas} K.~D.,    {Kr{\"u}ger} C.,  2013,
  Phys. Rev. D, 88, 044052

\bibitem[\protect\citeauthoryear{{Fantina}, {Chamel}, {Pearson} \&
  {Goriely}}{{Fantina} et~al.}{2012}]{bsk1}
{Fantina} A.~F.,  {Chamel} N.,  {Pearson} J.~M.,    {Goriely} S.,  2012,
  Journal of Physics Conference Series, 342, 012003

\bibitem[\protect\citeauthoryear{{Fantina}, {Chamel}, {Pearson} \&
  {Goriely}}{{Fantina} et~al.}{2013}]{bsk3}
{Fantina} A.~F.,  {Chamel} N.,  {Pearson} J.~M.,    {Goriely} S.,  2013,
  Astron. Astrop., 559, A128

\bibitem[\protect\citeauthoryear{{Ferrari}, {Miniutti} \& {Pons}}{{Ferrari}
  et~al.}{2003}]{ferra}
{Ferrari} V.,  {Miniutti} G.,    {Pons} J.~A.,  2003, MNRAS, 342, 629

\bibitem[\protect\citeauthoryear{{Finn}}{{Finn}}{1988}]{finn}
{Finn} L.~S.,  1988, MNRAS, 232, 259

\bibitem[\protect\citeauthoryear{{Flanagan} \& {Racine}}{{Flanagan} \&
  {Racine}}{2007}]{2007PhRvD..75d4001F}
{Flanagan} {\'E}.~{\'E}.,  {Racine} {\'E}.,  2007, Phys. Rev. D, 75, 044001

\bibitem[\protect\citeauthoryear{{Gaertig} \& {Kokkotas}}{{Gaertig} \&
  {Kokkotas}}{2009}]{gk}
{Gaertig} E.,  {Kokkotas} K.~D.,  2009, Phys. Rev. D, 80, 064026

\bibitem[\protect\citeauthoryear{{Garc{\'{\i}}a}, {Davies}, {Jim{\'e}nez},
  {Ballot}, {Mathur}, {Salabert}, {Chaplin}, {Elsworth}, {R{\'e}gulo} \&
  {Turck-Chi{\`e}ze}}{{Garc{\'{\i}}a} et~al.}{2013}]{2013JPhCS.440a2040G}
{Garc{\'{\i}}a} R.~A.,  {Davies} G.~R.,  {Jim{\'e}nez} A.,  {Ballot} J.,
  {Mathur} S.,  {Salabert} D.,  {Chaplin} W.~J.,  {Elsworth} Y.,  {R{\'e}gulo}
  C.,    {Turck-Chi{\`e}ze} S.,  2013, Journal of Physics Conference Series,
  440, 012040

\bibitem[\protect\citeauthoryear{{Gnedin} \& {Yakovlev}}{{Gnedin} \&
  {Yakovlev}}{1995}]{gnedin}
{Gnedin} O.~Y.,  {Yakovlev} D.~G.,  1995, Nucl. Phys. A, 582, 697

\bibitem[\protect\citeauthoryear{{Goriely}, {Chamel} \& {Pearson}}{{Goriely}
  et~al.}{2013}]{bsk2}
{Goriely} S.,  {Chamel} N.,    {Pearson} J.~M.,  2013, Phys. Rev. C, 88, 024308

\bibitem[\protect\citeauthoryear{{Gualtieri}, {Kantor}, {Gusakov} \&
  {Chugunov}}{{Gualtieri} et~al.}{2014}]{2014PhRvD..90b4010G}
{Gualtieri} L.,  {Kantor} E.~M.,  {Gusakov} M.~E.,    {Chugunov} A.~I.,  2014,
  \prd, 90, 024010

\bibitem[\protect\citeauthoryear{{Gusakov} \& {Andersson}}{{Gusakov} \&
  {Andersson}}{2006}]{gusa}
{Gusakov} M.~E.,  {Andersson} N.,  2006, MNRAS, 372, 1776

\bibitem[\protect\citeauthoryear{{Gusakov} \& {Haensel}}{{Gusakov} \&
  {Haensel}}{2005}]{gushae}
{Gusakov} M.~E.,  {Haensel} P.,  2005, Nucl. Phys. A, 761, 333

\bibitem[\protect\citeauthoryear{{Gusakov}, {Kaminker}, {Yakovlev} \&
  {Gnedin}}{{Gusakov} et~al.}{2004}]{2004A&A...423.1063G}
{Gusakov} M.~E.,  {Kaminker} A.~D.,  {Yakovlev} D.~G.,    {Gnedin} O.~Y.,
  2004, Astron. Astrop., 423, 1063

\bibitem[\protect\citeauthoryear{{Gusakov} \& {Kantor}}{{Gusakov} \&
  {Kantor}}{2013}]{gusk1}
{Gusakov} M.~E.,  {Kantor} E.~M.,  2013, Phys. Rev. D, 88, 101302

\bibitem[\protect\citeauthoryear{{Ho}, {Andersson} \& {Haskell}}{{Ho}
  et~al.}{2011}]{ho2011}
{Ho} W.~C.~G.,  {Andersson} N.,    {Haskell} B.,  2011, Phys. Rev. Lett., 107,
  101101

\bibitem[\protect\citeauthoryear{{Ho}, {Elshamouty}, {Heinke} \&
  {Potekhin}}{{Ho} et~al.}{2015}]{hoetal15}
{Ho} W.~C.~G.,  {Elshamouty} K.~G.,  {Heinke} C.~O.,    {Potekhin} A.~Y.,
  2015, Phys. Rev. C, 91, 015806

\bibitem[\protect\citeauthoryear{{Ho}, {Glampedakis} \& {Andersson}}{{Ho}
  et~al.}{2012}]{hoetal12}
{Ho} W.~C.~G.,  {Glampedakis} K.,    {Andersson} N.,  2012, MNRAS, 422, 2632; Erratum: 2012, MNRAS, 425,1600

\bibitem[\protect\citeauthoryear{{Kantor} \& {Gusakov}}{{Kantor} \&
  {Gusakov}}{2011}]{2011PhRvD..83j3008K}
{Kantor} E.~M.,  {Gusakov} M.~E.,  2011, Phys. Rev. D, 83, 103008

\bibitem[\protect\citeauthoryear{{Kantor} \& {Gusakov}}{{Kantor} \&
  {Gusakov}}{2014}]{gusk2}
{Kantor} E.~M.,  {Gusakov} M.~E.,  2014, MNRAS, 442, L90

\bibitem[\protect\citeauthoryear{{Kokkotas} \& {Sch\"afer}}{{Kokkotas} \&
  {Sch\"afer}}{1995}]{1995MNRAS.275..301K}
{Kokkotas} K.~D.,  {Sch\"afer} G.,  1995, MNRAS, 275, 301

\bibitem[\protect\citeauthoryear{{Kr{\"u}ger}, {Ho} \&
  {Andersson}}{{Kr{\"u}ger} et~al.}{2014}]{chris}
{Kr{\"u}ger} C.~J.,  {Ho} W.~C.~G.,    {Andersson} N.,  2014, preprint
  arXiv:1402.5656

\bibitem[\protect\citeauthoryear{{Lai}}{{Lai}}{1999}]{1999MNRAS.307.1001L}
{Lai} D.,  1999, MNRAS, 307, 1001

\bibitem[\protect\citeauthoryear{{Lee}}{{Lee}}{1995}]{1995A&A...303..515L}
{Lee} U.,  1995, Astron. Astrop., 303, 515

\bibitem[\protect\citeauthoryear{{Lopez-Monsalvo} \&
  {Andersson}}{{Lopez-Monsalvo} \& {Andersson}}{2011}]{2011RSPSA.467..738L}
{Lopez-Monsalvo} C.~S.,  {Andersson} N.,  2011, Royal Society of London
  Proceedings Series A, 467, 738

\bibitem[\protect\citeauthoryear{{McDermott}, {van Horn} \&
  {Hansen}}{{McDermott} et~al.}{1988}]{mcd}
{McDermott} P.~N.,  {van Horn} H.~M.,    {Hansen} C.~J.,  1988, Ap. J., 325,
  725

\bibitem[\protect\citeauthoryear{{Mendell}}{{Mendell}}{1991}]{mendell}
{Mendell} G.,  1991, Ap. J., 380, 530

\bibitem[\protect\citeauthoryear{{Miniutti}, {Pons}, {Berti}, {Gualtieri} \&
  {Ferrari}}{{Miniutti} et~al.}{2003}]{mini}
{Miniutti} G.,  {Pons} J.~A.,  {Berti} E.,  {Gualtieri} L.,    {Ferrari} V.,
  2003, MNRAS, 338, 389

\bibitem[\protect\citeauthoryear{{Ott}, {Burrows}, {Dessart} \& {Livne}}{{Ott}
  et~al.}{2006}]{ott}
{Ott} C.~D.,  {Burrows} A.,  {Dessart} L.,    {Livne} E.,  2006, Phys. Rev.
  Lett., 96, 201102

\bibitem[\protect\citeauthoryear{{Page}, {Geppert} \& {Weber}}{{Page}
  et~al.}{2006}]{2006NuPhA.777..497P}
{Page} D.,  {Geppert} U.,    {Weber} F.,  2006, Nucl. Phys. A, 777, 497

\bibitem[\protect\citeauthoryear{{Passamonti}, {Gaertig}, {Kokkotas} \&
  {Doneva}}{{Passamonti} et~al.}{2013}]{2013PhRvD..87h4010P}
{Passamonti} A.,  {Gaertig} E.,  {Kokkotas} K.~D.,    {Doneva} D.,  2013, Phys.
  Rev. D, 87, 084010

\bibitem[\protect\citeauthoryear{{Passamonti}, {Haskell}, {Andersson}, {Jones}
  \& {Hawke}}{{Passamonti} et~al.}{2009}]{pass1}
{Passamonti} A.,  {Haskell} B.,  {Andersson} N.,  {Jones} D.~I.,    {Hawke} I.,
   2009, MNRAS, 394, 730

\bibitem[\protect\citeauthoryear{{Potekhin}, {Fantina}, {Chamel}, {Pearson} \&
  {Goriely}}{{Potekhin} et~al.}{2013}]{pote}
{Potekhin} A.~Y.,  {Fantina} A.~F.,  {Chamel} N.,  {Pearson} J.~M.,
  {Goriely} S.,  2013, Astron. Astrop., 560, A48

\bibitem[\protect\citeauthoryear{{Prakash}, {Bombaci}, {Prakash}, {Ellis},
  {Lattimer} \& {Knorren}}{{Prakash} et~al.}{1997}]{prakash}
{Prakash} M.,  {Bombaci} I.,  {Prakash} M.,  {Ellis} P.~J.,  {Lattimer} J.~M.,
    {Knorren} R.,  1997, Phys. Reports, 280, 1

\bibitem[\protect\citeauthoryear{{Prix}, {Comer} \& {Andersson}}{{Prix}
  et~al.}{2002}]{prix}
{Prix} R.,  {Comer} G.~L.,    {Andersson} N.,  2002, Astron. Astrop., 381, 178

\bibitem[\protect\citeauthoryear{{Reisenegger} \& {Goldreich}}{{Reisenegger} \&
  {Goldreich}}{1992}]{rg}
{Reisenegger} A.,  {Goldreich} P.,  1992, Ap. J., 395, 240

\bibitem[\protect\citeauthoryear{{Weinberg}, {Arras} \& {Burkart}}{{Weinberg}
  et~al.}{2013}]{2013ApJ...769..121W}
{Weinberg} N.~N.,  {Arras} P.,    {Burkart} J.,  2013, Ap. J., 769, 121

\bibitem[\protect\citeauthoryear{{Yakovlev} \& {Pethick}}{{Yakovlev} \&
  {Pethick}}{2004}]{2004ARA&A..42..169Y}
{Yakovlev} D.~G.,  {Pethick} C.~J.,  2004, Ann. Rev. Astron. Astrophys., 42,
  169

\end{thebibliography}

%%%%%%%%%%%%%%%%%%%%%%%%%%%%%%%%%  LAST PAGE %%%%%%%%%%%%%%%%%%%%%%%%%%%%%%%
\label{lastpage}

%%%%%%%%%%%%%%%%%%%%%%%%%%%%%%%%%% END %%%%%%%%%%%%%%%%%%%%%%%%%%%%%%%%%%%%%%%%%%%%%%
\end{document}